\documentclass{aa}  

\usepackage{textcomp}
\usepackage{graphicx}
\usepackage{xcolor}
\usepackage{txfonts}
\usepackage{hyperref}
\usepackage{gensymb}
\usepackage[labelfont=bf]{caption}
\usepackage{subcaption}
\usepackage[labelfont=bf]{caption}
\usepackage[font={small}]{caption}
\usepackage[rightcaption]{sidecap}

\begin{document} 

   \title{Constraining the magnetic field in the galaxy cluster Abell 2142 using MeerKAT L-band polarisation data}

   \authorrunning{A. Pagliotta et al.}
   \titlerunning{A MeerKAT Rotation Measure study of Abell 2142}

   \author{A. Pagliotta
          \inst{1}\fnmsep\inst{2},
          C. J. Riseley\inst{3}\fnmsep\inst{1}\fnmsep\inst{2},
          A. Bonafede\inst{1}\fnmsep\inst{2},
          C. Stuardi\inst{2}
          \and 
          F. Loi\inst{4}}
   \institute{Dipartimento di Fisica e Astronomia, Università di Bologna, via Piero Gobetti 93/2, I-40129 Bologna, Italy\\
              \email{annalisa.pagliotta@unibo.it}
         \and 
         INAF - Istituto di Radioastronomia di Bologna, Via Piero Gobetti 101, I-40129 Bologna, Italy
         \and
         Astronomisches Institut der Ruhr-Universit\"{a}t Bochum (AIRUB), Universit\"{a}tsstra{\ss}e 150, 44801 Bochum, Germany
         \and INAF - Osservatorio Astronomico di Cagliari, Via della Scienza 5, Selargius, Italy
         }

   \date{Received Month Date, Year; accepted Month Date, Year; in original form Month Date, Year}
 
  \abstract
   {Magnetic fields permeate the Universe, including galaxy clusters, and affect the thermodynamical properties of the intra-cluster medium (ICM). Cosmological simulations predict that structure formation and mergers are capable of amplifying seed magnetic fields up to the $\mu$G level in the ICM, but the magnetic field strength and structure have only been studied in a few clusters.}
   {Abell 2142 is a local massive warm-cool-core cluster that shows evidence of a post-merger dynamical state. It hosts a multi-component radio halo and features both embedded and background polarised radio galaxies. In this work, we aim to constrain the magnetic field intensity, radial profile, and power spectrum within its ICM, providing key insights into the nature of this non-thermal component in galaxy clusters.}
   {We present MeerKAT observations of Abell 2142 in the L-band (872-1712 MHz), imaged in polarisation for the first time with this purpose. We derived the rotation measure (RM) from the polarised emission of radio galaxies by applying the RM synthesis technique and analyse both the RM and fractional polarisation ($\mathrm{F_p}$). To investigate the magnetic field distribution within the ICM, we compared our results with mock RM maps generated from 3D simulations of the cluster.}
   {We find that the RM dispersion, $\sigma_{\mathrm{RM}}$, decreases with the projected radius, whereas the $\mathrm{F_p}$ increases. Both trends suggest that the magnetic field intensity decreases at larger distances from the cluster centre, in agreement with studies of other clusters. Assuming  the magnetic field energy density scales with the gas thermal energy ($\rm B \propto n_e^{0.5}$), we find the best fit to our data is a magnetic field with a power spectrum ranging between 7 and 470 kpc, with peak at $\sim 140$ kpc and mean central strength of $9.5 \pm 1.0 \ \mu$G.}
   {The high central magnetic field lies at the upper end of the range observed in other systems and supports the possibility of an hadronic contribution to the diffuse radio emission previously detected at the cluster centre.}

   \keywords{magnetic fields; galaxy cluster: individual: Abell 2142; non-thermal radio emission; polarisation; extragalactic astrophysics.}
   
   \maketitle

\section{Introduction}

Magnetic fields (MFs) are ubiquitous in the Universe and it is broadly accepted that they permeate the volume of galaxy clusters up to Mpc scales (see \citealt{Govoni_2004} and \citealt{Ferretti_2012}). The study of this non-thermal component is relevant to understand the dynamics and energetics in cluster environments, since MFs provide an additional term of pressure and influence the thermal conduction (e.g. \citealt{Malyshkin_2001}; \citealt{Roberg_2016}). Moreover, they play an important role in the acceleration of particles, diffusion of cosmic rays (CRs), and formation of diffuse radio large-scale structures, known as radio halos, relics, and mini-halos (see \citealt{VanWeeren_2019}). 

The origin of MFs in galaxy clusters is still debated, as they lose the `memory' of the initial seeds due to a small-scale dynamo amplification; however, recent measurements of the MF in filaments have strongly suggested a primordial origin (see e.g. \citealt{Carretti_2022}, \citeyear{Carretti_2023}). Moreover, MFs are well constrained thanks to state-of-the-art techniques in only a few clusters: therefore, very little is known about their amplification and evolution (see \citealt{Ryu_2012} and \citealt{Donnert_2018}). Cosmological simulations predict that structure formation, accretion, and merger-driven shocks and turbulence are able to amplify weak seed MFs through adiabatic compression and dynamos (e.g. \citealt{Ryu_2008}; \citealt{Iapichino_2012}; \citealt{Vazza_2014}, \citeyear{Vazza_2017}, \citeyear{Vazza_2018}; \citealt{DF_2019}; \citealt{Botteon_2022}). Through these mechanisms, MFs are allowed to reach the observed $\mu$G levels in local clusters, fluctuating over spatial scales of few to 100s kpc. Intensities from $5-10 \ \mu$G are measured in the central regions (e.g. \citealt{Murgia_2004}; \citealt{Bonafede_2010}; \citealt{Govoni_2017}; \citealt{Stuardi_2021}), with values exceeding 10 $\mu$G in the core of cool-core (CC) clusters (e.g. \citealt{Vacca_2012}) and decreasing down to fractions of $\mu$G in the periphery (e.g. \citealt{Bonafede_2013}). In addition, there is emerging evidence that high redshift galaxy clusters could have similar MF strengths to local clusters (e.g. \citealt{DiGennaro_2020}), implying that MFs are amplified rapidly during cluster formation.\\
\\
The most promising technique, which has been used thus far to derive information about the ICM MF, is the analysis of the ‘Faraday rotation effect' on radio galaxies located both inside and in the background of galaxy clusters. Radio galaxies produce partially linearly polarised synchrotron radiation from their jets and lobes. The polarisation angle, $\psi$, of these waves is rotated by a quantity proportional to the squared observing wavelength, $\lambda^2$, and to the ‘rotation measure' (RM) once it passes through the ICM (i.e. an ionised and magnetised plasma; e.g. \citealt{Burn_1966}). The latter term is directly linked to the properties of the rotating region and defined as
\begin{equation}
    \rm RM = 812 \int_{source}^{observer} n_e B_{\parallel} \mathrm{d}l \ \ \ \mathrm{rad \ m^{-2}},
    \label{eq:RM}
\end{equation}
where $\rm n_e$ is the number density of free electrons in $\mathrm{cm^{-3}}$, $\rm B_{\parallel}$ is the component of the MF along the line of sight (LOS) in $\mu$G, and dl is the infinitesimal path along the LOS in kpc. The RM is conventionally assumed to be positive or negative depending on the orientation of the MF along the LOS (i.e. towards or away from the observer, respectively). More generally, it is referred to as the rotation of the angle, $\psi$, originating from an infinitesimal volume, with LOS length d$\vec{\textbf{r}}$, in which the medium can be decomposed, and is called the Faraday depth, $\phi$. The RM and the Faraday depth coincide at all wavelengths only when one or several non-emitting regions lie between the source and the observer.
More complex scenarios also affect the polarisation fraction of the sources and should be treated with advanced techniques, such as the ‘RM synthesis technique' \citep{Brentjens_2005}. \par The study of the average RM, $\langle \mathrm{RM} \rangle$, its dispersion, $\sigma_{\mathrm{RM}}$, and the fractional polarisation, $\mathrm{F_p}$, extracted from radio sources at different projected distances from the cluster centre, allows us to get the first hints on the distribution of the LOS MF across the ICM. To constrain the MF intensity and profile, the key approach is to compare the observed RM map with mock RM maps obtained from ad hoc simulations, which model a 3D MF from a specific power spectrum. This power spectrum is not fully known, but recent cosmological magneto-hydrodynamical (MHD) simulations accounting for small-scale dynamo amplification (\citealt{DF_2019}) have revealed that it is more complex than the simple power law, such as a Kolmogorov power spectrum, that was used in the first works (see e.g. \citealt{Murgia_2004}; \citealt{Bonafede_2010}; \citealt{Vacca_2010}). \par In this paper, we performed a radio analysis of the polarisation of a local massive galaxy cluster, Abell 2142 (A2142) to constrain, for the first time, the MF intensity, profile, and power spectrum in its ICM. As discussed in Sect. \ref{Abell2142}, A2142 is located at the centre of the Abell 2142 supercluster, placing it in continuous interaction with the surrounding medium. Moreover, its intermediate dynamical state (i.e. between merging and relaxed) makes it a particularly intriguing environment for exploring and studying the MF.
We used high-sensitivity MeerKAT L-band (872-1712 MHz) data, which were previously processed and published in continuum by \citealt{Riseley_2024}; here, they have been imaged  for the first time in polarisation. In Sect. \ref{Data Analysis}, we describe briefly the imaging procedure, both in total and in polarised intensity, and the RM synthesis technique main steps. In Sect. \ref{Results}, we present the results from the data analysis, from which we derive the $\langle \mathrm{RM} \rangle$, $\sigma_{\mathrm{RM}}$, and $\mathrm{F_p}$ projected radial profiles. The simulations we  used are described in Sect. \ref{Simulations}, together with the statistical comparison with the observations and the constrain of the MF properties. Section \ref{Discussion} is dedicated to the discussion of the previous outcomes and in Sect. \ref{Conclusion} we present a brief summary of our work. \par Throughout the paper a $\Lambda$CDM cosmology is assumed, with $\rm H_0 = 70 \ \mathrm{km \ s^{-1} \ Mpc^{-1}}$, $\Omega_{\mathrm{M}} = 0.3,$ and $\Omega_{\Lambda} = 0.7$. This translates to a luminosity distance of $\rm D_L = 408.6$ Mpc and a scale of 1.669 kpc/arcsec at the cluster redshift, z = 0.0894.

\section{Abell 2142} \label{Abell2142}
\begin{figure*}
    \begin{subfigure}{0.569\linewidth}
        \hspace{-0.3cm}
        \includegraphics[width=0.95\linewidth]{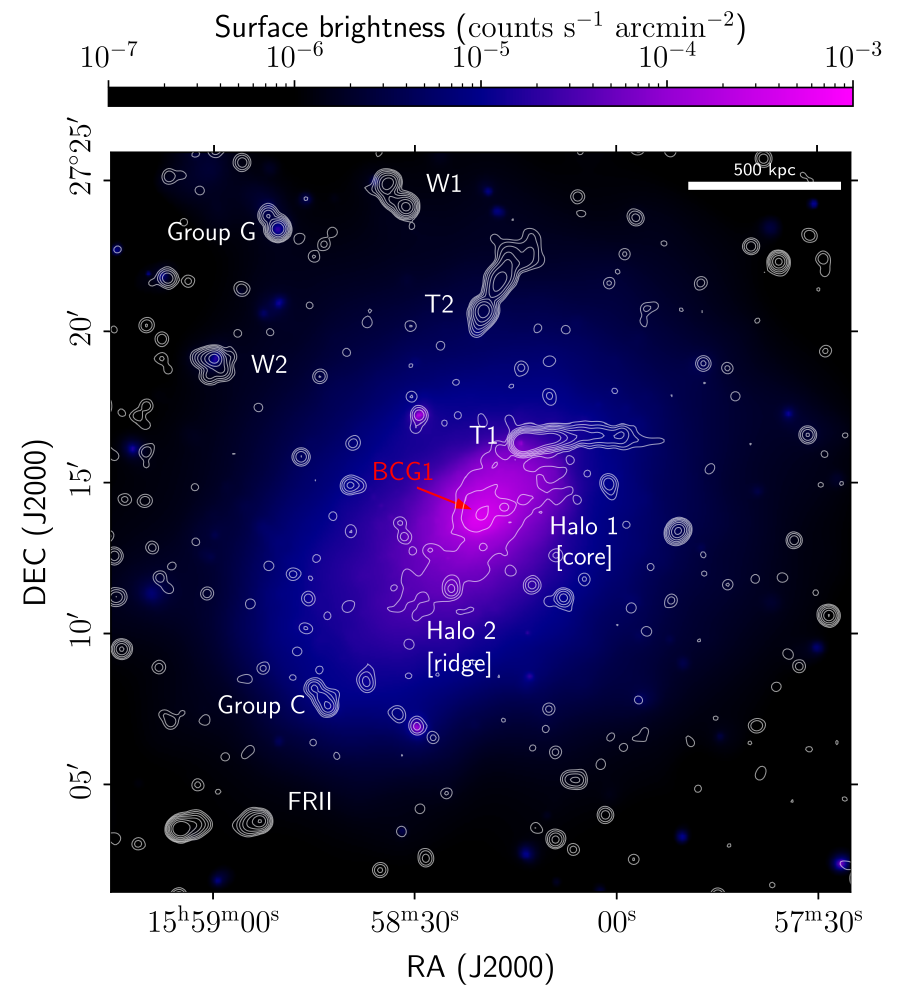}
        \caption{}
        \label{subfig:X-ray}
    \end{subfigure}
    \hspace{-0.8cm}
    \begin{subfigure}{0.545\linewidth}
        \hspace{-0.3cm}
        \includegraphics[width=0.95\linewidth]{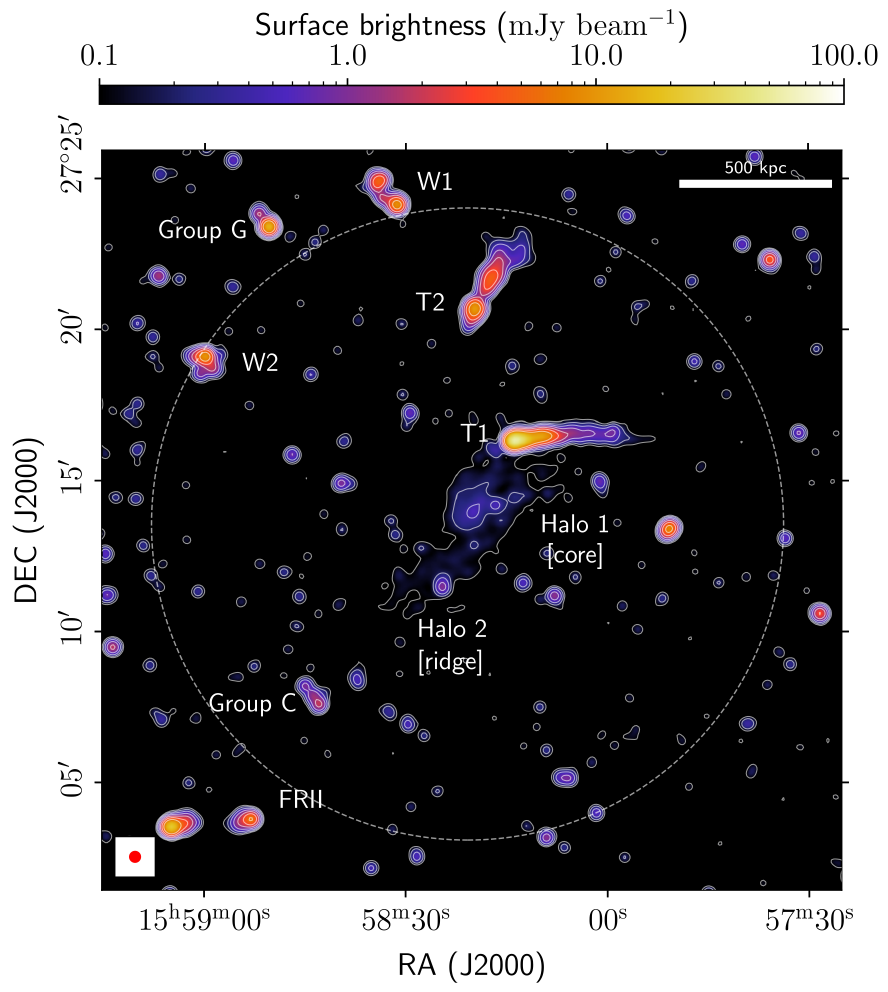}
        \caption{}
        \label{subfig:Stokes-I_zoom}
    \end{subfigure}
    \caption[]{Abell 2142. Radio contours at 1283 MHz, starting from $3\sigma_{\mathrm{I}}$ and scaling by a factor of 2, are overlaid in both maps. The known radio galaxies and two out of three halo components are labeled with their names. Only BCG1 is pointed in red in panel (a), as BCG2 does not show evidence of any radio emission. Panel (a): XMM Cluster Outskirts Project (X-COP) adaptively smoothed, vignetting-corrected and background-subtracted X-ray surface brightness in the range [$0.7 - 1.2$] keV\footnotemark. Panel (b): Total intensity (Stokes I) map of A2142 at 1283 MHz. It is appreciable the emission from both radio galaxies and the two radio halos, H1 and H2. Dashed circle corresponds to the reference distance of 1 Mpc from the cluster centre. The spatial scales and resolution beam of $20^{\prime\prime} \times 20^{\prime\prime}$ (33.4 kpc) are reported on the edges of the image.}
    \label{fig:A2142}
\end{figure*}

A2142 is a massive cluster found in the local Universe at redshift $z = 0.0894$ \citep{Boehringer_2000}. It is the dominant central member of the Abell 2142 galaxy supercluster, which spans up to 35 Mpc and comprises 950 galaxies within $\sim 3$ Mpc. They are hierarchically organised in many structures and substructures, associated with infalling small galaxies and groups (\citealt{Okabe_2008}). \citealt{Owers_2011}  spectroscopically confirmed this hierarchical configuration and found that most substructures account for $\sim 10$ members, indicating that there is no sign of recent or ongoing major mergers, but possibly minor events. The main properties of A2142 are reported in Table \ref{tab:A2142specifics}.
\begin{table}[!htb]
    \vspace{-0.3cm}
    \centering
    \caption[]{\label{tab:A2142specifics} Properties of A2142.}
    \begin{tabular}{lcccccccc}
        \hline 
        \noalign{\smallskip}
        \noalign{\smallskip}
        $\mathrm{RA_{J2000}}$ & & & 15h 58m 21s & & & (1) \\
        \noalign{\smallskip}
        $\mathrm{DEC_{J2000}}$ & & & 27° 13' 37'' & & & (1) \\
        \noalign{\smallskip}
        $z$ & & & 0.0894 & & & (2)\\
        \noalign{\smallskip}
        $\rm M_{500}$ & & & $(8.8 \pm 0.2) \times 10^{14} \ \mathrm{M_{\odot}}$ & & & (3) \\
        \noalign{{\smallskip}}
        $\rm R_{500}$ & & & $1408.5 \pm 70.4$ kpc & & & (3) \\
        \noalign{\smallskip}
        \hline
    \end{tabular}
    \tablefoot{$\rm M_{500}$ is the mass contained within $\rm R_{500}$ and $\rm R_{500}$ is the radius at which the density is 500 times the critical density of matter in the Universe at the cluster redshift.}
    \tablebib{(1): \citealt{Yuan_2022}; (2): \citealt{Boehringer_2000}; (3): \citealt{PC_2016}.}
\end{table}
\vspace{-0.2cm}

Studies carried out in the X-ray band using Chandra and XMM-Newton have demonstrated the non-relaxed nature of this object, that presents an asymmetrically distributed ICM (see Fig. \ref{subfig:X-ray}), extending along the north-west/south-east axis, aligned with the filamentary structure of the supercluster \citep{Einasto_2015}.
Four spiral-like cold fronts have been observed as a direct consequence of a core sloshing episode, with three of them on small scales ($\sim$ kpc). The fourth one is located on larger scales, $\sim 1$ Mpc, to the south-east side, for which the origin is still unclear (\citealt{Markevitch_2000}; \citealt{Rossetti_2013}). A2142 is classified as a warm-cool-core cluster as it shares properties of both relaxed and merging systems \citep{Wang_2018}. In this case, a well-defined X-ray peak is present as in relaxed structures, with a central electron density of $\rm n_{e,0} = 1.8\times10^{-2} \ \mathrm{cm^{-3}}$ \citep{Tchernin_2016}, but the temperature and entropy are not consistent with typical values found for CC clusters: the gas is cooler ($\lesssim$ 7 keV) at the centre with respect to the global temperature, $\sim 8-9$ keV \citep{Tchernin_2016}, and the central entropy, $\rm K_0$, is of the order of $49 \ \mathrm{keV \ cm^{2}}$ \citep{Wang_2018}. \citealt{Markevitch_2000} suggested that a past intermediate-mass-ratio merger, viewed at a time at least $1-2$ Gyr after the initial core crossing, was not able to disrupt entirely the cool-core. Thus, together with subsequent minor mergers, it could explain such features and the observed cold fronts \citep{Wang_2018}. 

Further evidence for this scenario comes from the optical side of the electromagnetic spectrum revealing the presence of two faint brightest cluster galaxies (BCGs), BCG1 and BCG2. BCG1 ($z_{\rm spec} = 0.0908$) at the centre of the gravitational potential (see Fig. \ref{subfig:X-ray}), close to the innermost cold front, with an offset of $\approx 30$ kpc from the gas density peak (\citealt{Wang_2018}). BCG2 ($z_{\rm spec} = 0.0965$) is, on the other hand, located to the north-west of the cluster centre, possibly belonging to another merging group and having a relative velocity of $\sim 1800 \ \mathrm{km \ s^{-1}}$ \citep{Oegerle_1995}. 

Moving to the radio band, A2142 shows an interesting multi-component radio halo, studied in \citealt{Venturi_2017}, \citealt{Bruno_2023}, and \citealt{Riseley_2024}. The central component (‘Halo 1’, H1, or ‘core’; see Fig. \ref{subfig:Stokes-I_zoom}) is more roundish and centrally located; whereas the second component (‘Halo 2’, H2, or ‘ridge’; see Fig. \ref{subfig:Stokes-I_zoom}) is more ridge-like and extends to the south-east direction.
\citealt{Riseley_2024} confirmed the presence of a third halo component (‘Halo 3’, H3), first reported by \citealt{Bruno_2023}, with an elliptical morphology elongated in the north-west and south-east direction, encompassing the other structures and filling a volume out to $\sim 1.2$ Mpc from the cluster centre. This large-scale component is not visible in Fig. \ref{subfig:Stokes-I_zoom}.
All the components of the halo have steep radio synchrotron spectra ($\rm S_{\nu} \propto \nu^{-\alpha}$, where $\alpha$ is the spectral index), $\alpha_{\mathrm{H1}} = 1.09 \pm 0.03$, $\alpha_{\mathrm{H2}} = 1.15 \pm 0.04$ and H3 showing an ultra-steep radio spectrum with $\alpha_{\mathrm{H3}} = 1.68 \pm 0.10$. 
Both \citealt{Bruno_2023} and \citealt{Riseley_2024} found a sub-linear correlation between the radio surface brightness and the X-ray surface brightness for H1, suggesting that the non-thermal component is much broader than the thermal one. \citealt{Bruno_2023} discussed the scenario in which secondary CR electrons contribute to the emission of H1: in this case, the sub-linear value of the index $\rm k$ on scales of $\sim 100 - 200$ kpc would indicate a flat CR proton distribution in the core and a strong MF ($\rm B > 4 \ \mu$G).

Finally, several radio galaxies, cluster members and background sources, populate the field of view  (see Fig. \ref{subfig:Stokes-I_zoom} and Table \ref{tab:radiogspecifics}). Following the nomenclature used by \citealt{Venturi_2017} and \citealt{Bruno_2023}, we named them: two head-tail radio galaxies, T1 and T2, a defined Fanaroff-Riley II (FRII) galaxy, 7C 1557+2712, located in the south-east side of the cluster, two wide-angle-tails (WATs), W1, and W2, along with the galaxy groups G (a blend of discrete radio sources located in the north-east to the cluster centre), and C (in the south-east).

\footnotetext{available at \url{https://dominiqueeckert.wixsite.com/xcop}.}

\begin{table}[!htb]
    \centering
    \caption[]{\label{tab:radiogspecifics}Known background and cluster member radio galaxies listed with their corresponding redshift.}
    \begin{tabular}{lcccccccccccc}
        \hline 
        \noalign{\smallskip}
        \noalign{\smallskip}
        T1 & & & $z_{\rm spec} = 0.0954$ & & & M & & & (1) \\
        \noalign{\smallskip}
        T2 & & & $z_{\rm spec} = 0.0895$ & & & M & & & (1) \\
        \noalign{\smallskip}
        W1 & & & $z_{\rm phot} = 0.5740$ & & & B & & & (2)\\
        \noalign{\smallskip}
        W2 & & & / & & & B & & & (*) \\
        \noalign{\smallskip}
        Group G & & & $z_{\rm spec} = 0.0940$ & & & B & & & (3) \\
        \noalign{\smallskip}
        Group C & & & / & & & M & & & (*)\\
        \noalign{\smallskip}
        FR II  & & & $z_{\rm phot} = 0.8560$ & & & B & & & (4) \\
        \noalign{\smallskip}
        \hline
    \end{tabular}
    \tablefoot{M: cluster member; B: background; (*) for these sources the redshift is not known, but according to \citealt{Venturi_2017}, W2 is not a cluster member, whereas Group C is associated with other cluster galaxies with similar optical magnitude.}
    \tablebib{(1): \citealt{Ahumada_2020}; (2): \citealt{Venturi_2017}; (3): \citealt{Eckert_2014}; (4): \citealt{Duncan_2022}.}
\end{table}

\section{Data analysis} \label{Data Analysis}
\subsection{Calibration and imaging} 
A2142 was observed with the MeerKAT radio interferometer in the L-band frequency range ($872-1712$ MHz) under the Project ID (PID) SCI-20210212-CR-01 (P.I. Riseley). Due to the high declination of A2142, \citealt{Riseley_2024} performed two observations (10. October and 12. November 2021) to achieve the target sensitivity, resulting in a total on-source time of 5.5 hours.

For the calibration, \citealt{Riseley_2024} adopted the Containerized Automated Radio Astronomy Calibration (\texttt{CARACal}) pipeline \citep{Jozsa_2020}, which employs calibration tasks from the Common Astronomy Software Application (\texttt{CASA}) package. We refer to \citealt{Riseley_2024} for specifics on calibration of these data and \citealt{Loi_2025} for details on the pipeline development and strategies.\\
\\
To apply the RM synthesis technique (see Sect. \ref{RMS Technique}), images of the Stokes Q and Stokes U polarisations of a measurement set are required at different frequencies to create the corresponding cubes. Both the images in total intensity and in polarisation (Stokes I, Q, U and V\footnote{We produced the Stokes V images as a verification check on our calibration, since measured degrees of Stokes V usually range between $0.1\%$ and $0.5\%$ of the Stokes I intensity (\citealt{Ruszkowski_2002} and references therein).}) were obtained with the \texttt{WSClean v3.4.0} (‘\emph{w}-Stacking Clean’) software\footnote{Available at \url{https://gitlab.com/aroffringa/wsclean}.} (\citealt{Offringa_2014}, \citeyear{Offringa_2017}). 
We performed a multi-frequency joint polarimetric deconvolution dividing the full-bandwidth in 256 channels (half the number of channels used in the reduction step). This was done both to limit the computational load and to verify the maximum Faraday depth, $|\phi_{\rm max}|$, at which we are sensitive, given this configuration (see Sect. \ref{RMS Technique}). We used a Briggs weighting \citep{Briggs_1995} with a robust parameter equal to $- 0.5$, in order to recover both small scale and extended emission in the observations and then corrected for the primary-beam of MeerKAT antennas\footnote{Other instruments are present in the EveryBeam Library, available at \url{https://everybeam.readthedocs.io/en/latest}.}. We  used the multi-scale deconvolution option, using scales between 0 (point sources) and 10, in units of pixels, and applied a mask in order to perform a deeper cleaning down to the $1.5\sigma$ threshold. 

\subsection{Image preparation for the RM synthesis} \label{IMprep}
Before starting with the RM synthesis technique, the sub-band Stokes Q and U images must be convolved to the same angular resolution to create the cubes in frequency. We opted for a circular beam of $20^{\prime\prime} \times 20^{\prime\prime}$ (33.4 kpc) and smoothed the images with the \texttt{imsmooth CASA} task. A few channels with coarser resolution beam due to flagging and high noise levels were discarded. The resulting Stokes Q and U cubes cover a full-bandwidth between 884 to 1670 MHz, divided in 191 usable channels with width of 3.34 MHz. Finally, the cubes were re-binned for computational load to the desired pixel scale of 4$^{\prime\prime}$ ($\sim 6.7$ kpc). We show in Fig. \ref{subfig:Stokes-I_zoom} the Stokes I multi-frequency-synthesis (MFS) image after the convolution and the re-binning.

\subsection{RM synthesis technique} \label{RMS Technique}
To recover the polarised emission at multiple Faraday depths along a particular LOS and to derive the RM, we used the RM synthesis technique introduced by \citealt{Brentjens_2005}. We computed the resolution in Faraday space\footnote{$\delta\phi$ is also the Full-Width-Half-Maximum (FWHM) of the Rotation-Measure-Transfer-Function (RMTF).}, $\delta\phi$, the maximum observable $\phi$, $|\phi_{\mathrm{\rm max}}|$, and the maximum observable scale in Faraday space, $\Delta\phi_{\rm max}$, of our L-band observations (see \citealt{Brentjens_2005}). Moreover, we checked for the theoretical noise in RM, $\mathrm{err_{RM}}$, assuming a signal-to-noise (S/N) ratio of 6\footnote{which corresponds to a Gaussian significance level of about $4.8\sigma$ according to \citealt{Hales_2012}.} (see \citealt{SB_2013}). All these parameters are summarised for our case in Table \ref{tab:RMSparameters}.
\begin{table}[!htb]
    \centering
    \caption[]{\label{tab:RMSparameters}List of parameters for the RM synthesis technique.}
    \begin{tabular}{ccccccc}
        \hline
        \noalign{\smallskip}
        \noalign{\smallskip}
        $\delta\phi$ & &$|\phi_{\rm max}|$ & & $\Delta\phi_{\rm max}$ & & $\mathrm{err}_{\mathrm{RM}}$ \\
        $\mathrm{[rad \ m^{-2}]}$ & & $\mathrm{[rad \ m^{-2}]}$ & & $\mathrm{[rad \ m^{-2}]}$ & & $\mathrm{[rad \ m^{-2}]}$ \\
        (1) & & (2) & & (3) & & (4) \\
        \noalign{\smallskip}
        \hline
        \noalign{\smallskip}
        \noalign{\smallskip}
        41.82 & & 2004.18 & & 97.84 & & 3.48\\
        \noalign{\smallskip}
        \hline
    \end{tabular}  
    \tablefoot{(1) theoretical FWHM of the RMTF; (2) maximum observable $\phi$; (3) maximum observable scale in Faraday space; (4) theoretical noise in RM at S/N = 6.}
\end{table}
\\
With this data configuration, we were able to detect typical values of the RM observed in other cluster of galaxies (see e.g. \citealt{Boehringer_2016}). We applied the RM synthesis technique running the \texttt{rmsynth3d} task implemented in the Canadian Initiative for Radio Astronomy Data Analysis \textsc{(CIRADA) RM-Tools}\footnote{available at \url{https://github.com/CIRADA-Tools/RM-Tools}.}, which takes as input the Stokes Q and U cubes with the associated frequency list. We therefore obtained the reconstructed total, $\tilde{F}(\phi)$, real, $\tilde{Q}(\phi)$, and imaginary, $\tilde{U}(\phi)$, cubes of the dirty Faraday dispersion function (FDF) in Faraday space, ranging between $\pm \ 999 \ \mathrm{rad \ m^{-2}}$ in steps of $3 \ \mathrm{rad \ m^{-2}}$ (667 channels). We halved the maximum observable $\phi$ to reduce the computational load and, since $|\phi_{\rm max}|$ is not a multiple of the sample spacing ($3 \ \mathrm{rad \ m^{-2}}$), the algorithm decreases it to the nearest multiple (i.e. 999 $\mathrm{rad \ m^{-2}}$).\\
\\
At this point, we performed the RM deconvolution (similar to the aperture synthesis imaging; see \citealt{Heald_2009}) with the \texttt{rmclean3d} task to remove some possible contamination by the RMTF sidelobes after the convolution. We determined the RMS noise level of the real and imaginary part of the dirty FDF from the edges of the cubes, $|\phi| > 750 \ \mathrm{rad \ m^{-2}}$, where there was no evidence of contamination by the sidelobes of the sources. The values are computed as the mean over the pixels contained in a large box without sources and located within the primary beam. We obtained:
\begin{itemize}
    \itemsep8pt
    \vspace{0.05cm}
    \item[-] $\sigma_{\mathrm{Q}} = 4.74 \ \mu\mathrm{Jy \ beam^{-1}}$ as RMS noise for the Stokes Q cube;
    \vspace{-0.2cm}
    \item[-] $\sigma_{\mathrm{U}} = 4.70 \ \mu\mathrm{Jy \ beam^{-1}}$ as RMS noise for the Stokes U cube;
    \item[-] $\sigma_{\mathrm{QU}} = \dfrac{\sigma_{\mathrm{Q}}+\sigma_{\mathrm{U}}}{2} = 4.72 \ \mu\mathrm{Jy \ beam^{-1}}$ as average RMS noise.
    \vspace{0.2cm}
\end{itemize}
Following \citealt{George_2012}, we used a conservative limit of 8 times $\sigma_{\mathrm{QU}}$ ($37.74 \ \mu\mathrm{Jy \ beam^{-1}}$) as the threshold for \texttt{rmclean3d}. This corresponds to a Gaussian significance level of about $7\sigma$ according to \citealt{Hales_2012}.

\subsection{Final maps}
From the cleaned FDF cubes we created the final maps for the further analysis on the RM and the $\mathrm{F_p}$ radial profiles. All the images have been obtained by masking for $6\sigma_{\mathrm{QU}}$ and three times the noise in Stokes I intensity ($\sigma_{\mathrm{I}} = 3.26 \times 10^{-5} \ \mathrm{Jy \ beam^{-1}}$). Compared to the previous cleaning step, we adopted a lower polarisation threshold to include a larger number of pixels, thereby minimising the risk of underestimating the RM values (since the sources with low polarised intensity can exhibit higher RM moduli). At the same time, we mitigated contamination from the noise by applying a mask in total intensity.  

\begin{figure}[!htb]
    \vspace{-0.3cm}
    \centering
    \includegraphics[width=1\linewidth]{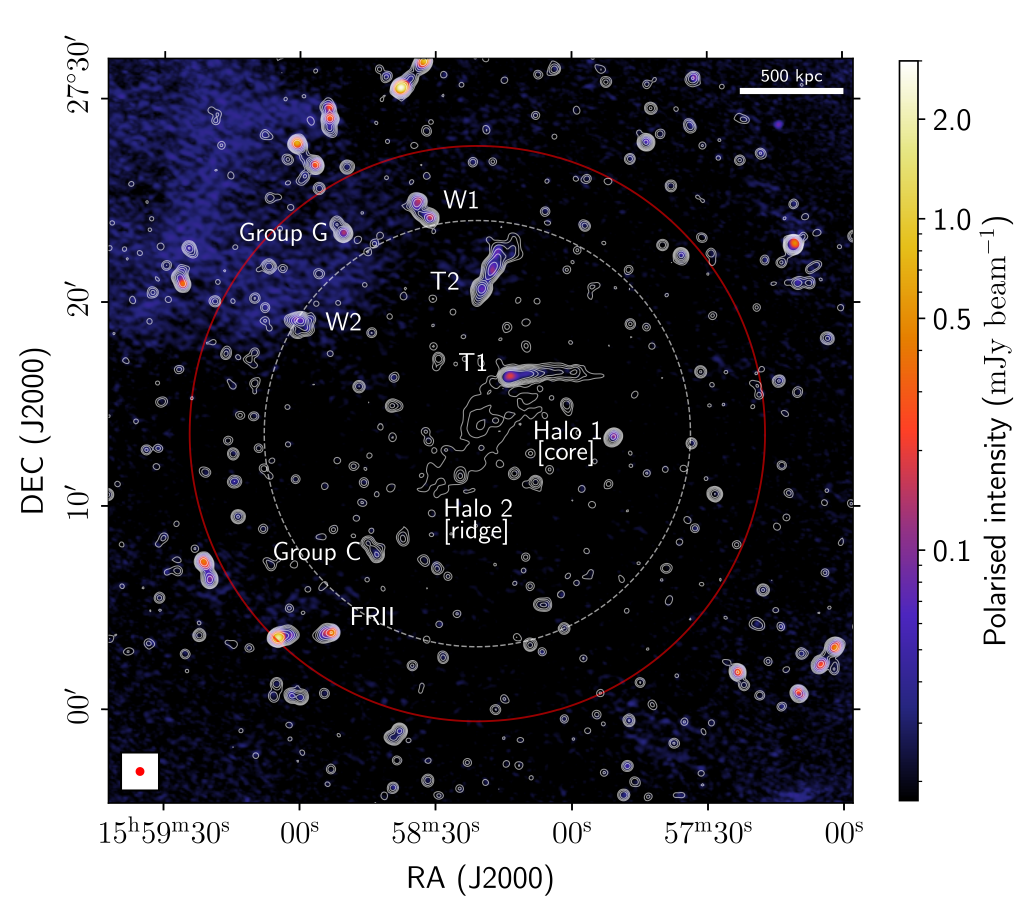}
    \caption{Map of the maximum polarised intensity in each pixel obtained with the RM synthesis technique, with overlaid radio contours from $3\sigma_{\mathrm{I}}$ and scaling by a factor of 2. The Galactic bubble polarised emission can be seen in the top-left side of the image. Dashed circle corresponds to the reference distance of 1 Mpc from the cluster centre. Solid red line represent the 0.23$^{\degree}$ ($\sim 1381$ kpc) separation threshold for MeerKAT off-axis leakage in the L-band above 1.4 GHz. Spatial scales and resolution beam are reported on the edges of the map.\\ }
    \label{fig:max-PI}
    \vspace{-1.2cm}
\end{figure}

The RM map was  created by taking the $\phi$ corresponding to the peaks of the cleaned FDF cube, $\phi_{\rm peaks}$, which represents the equivalent of the RM map in the case of ‘Faraday-simple' screens (i.e. characterised by a single value of $\phi$), and corrected for the Galactic Faraday rotation in the region of the cluster ($\mathrm{RM_{Gal}} = 4.10 \pm 1.13 \ \mathrm{rad \ m^{-2}}$; \citealt{Hutschenreuter_2022}). We  then produced the polarisation map with the maximum polarised intensity in each pixel from the cleaned FDF cube, $|\tilde{F}(\phi_{\rm peak})|$, (see Fig. \ref{fig:max-PI}) and corrected it for the Ricean bias \citep{George_2012}. For the $\mathrm{F_p}$ map, we  divided the obtained polarisation map by the convolved and re-binned total intensity MFS image.

\section{Results} \label{Results}
This section is dedicated to the presentation of the results obtained with the RM synthesis technique and the subsequent analysis of the RM and $\mathrm{F_p}$ maps (Sect. \ref{RM Synthesis Results}). The first considerations on the MF properties and distribution in the ICM of A2142 are instead illustrated in Sect. \ref{Radial profiles} from the investigation of the projected radial profiles for some sources in the field.

\subsection{RM synthesis results} \label{RM Synthesis Results}
As mentioned above, the RM synthesis technique delivers three cubes of the dirty FDF spectrum. From the one containing the amplitude (i.e. the polarised intensity), the algorithm can produce an additional map with the maximum polarised intensity in each pixel (i.e. the peak of the FDF spectrum, $|\tilde{F}(\phi_{\rm peak})|$, in each pixel). We show this image in Fig. \ref{fig:max-PI}, where numerous polarised sources, including some associated with A2142 and several field galaxies, are clearly visible. On the one hand, the two radio halo components, H1 and H2, are heavily depolarised. 

\begin{figure}[!htb]
    \begin{subfigure}{1\linewidth}
        \centering
        \includegraphics[width=0.9\linewidth]{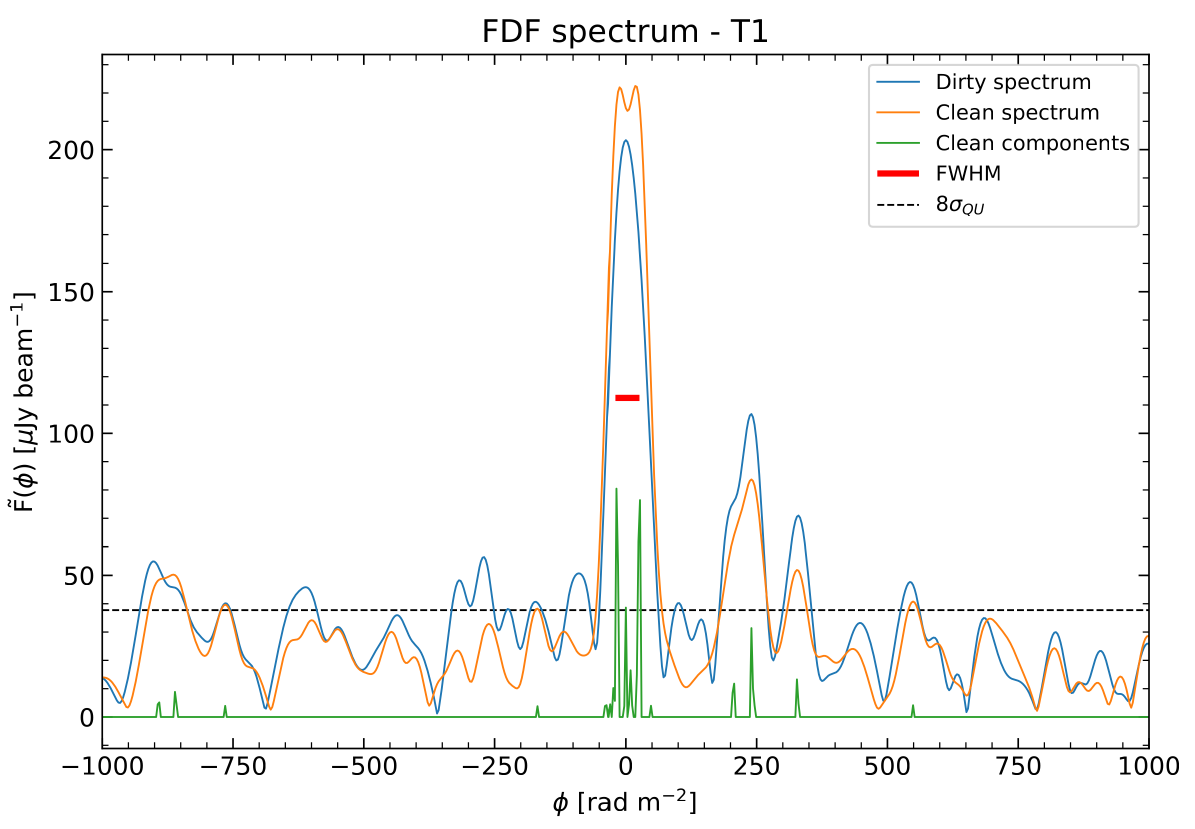}
        \caption{}
        \label{subfig:FDF-T1}
    \end{subfigure}
    \begin{subfigure}{1\linewidth}
        \vspace{0.2cm}
        \centering
        \includegraphics[width=0.9\linewidth]{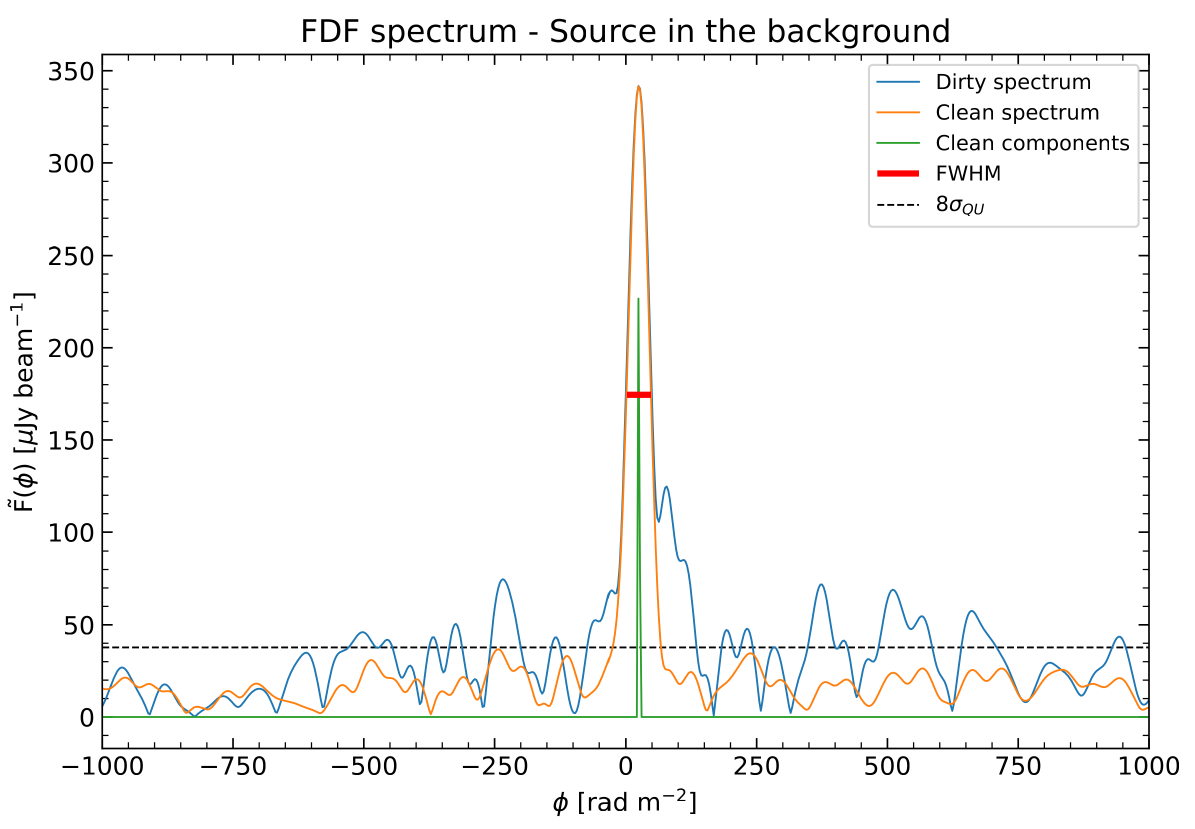}
        \caption{}
        \label{subfig:FDF-Sbkg}
    \end{subfigure}
    \caption{Reconstructed FDF spectrum taken from the brightest polarised pixel of some sources in the field. Panel (a): T1. Panel (b): Source in the background. The dirty spectrum is shown in blue, the clean spectrum in orange and the clean components in green. The FWHM of the RMTF is reported in red and the $8\sigma_\mathrm{{QU}}$ with the black dashed line as a reference.}
    \label{FDF}
\end{figure}
\begin{figure*}
    \begin{subfigure}{0.52\linewidth}
        \hspace{-0.2cm}
        \includegraphics[width=0.91\linewidth]{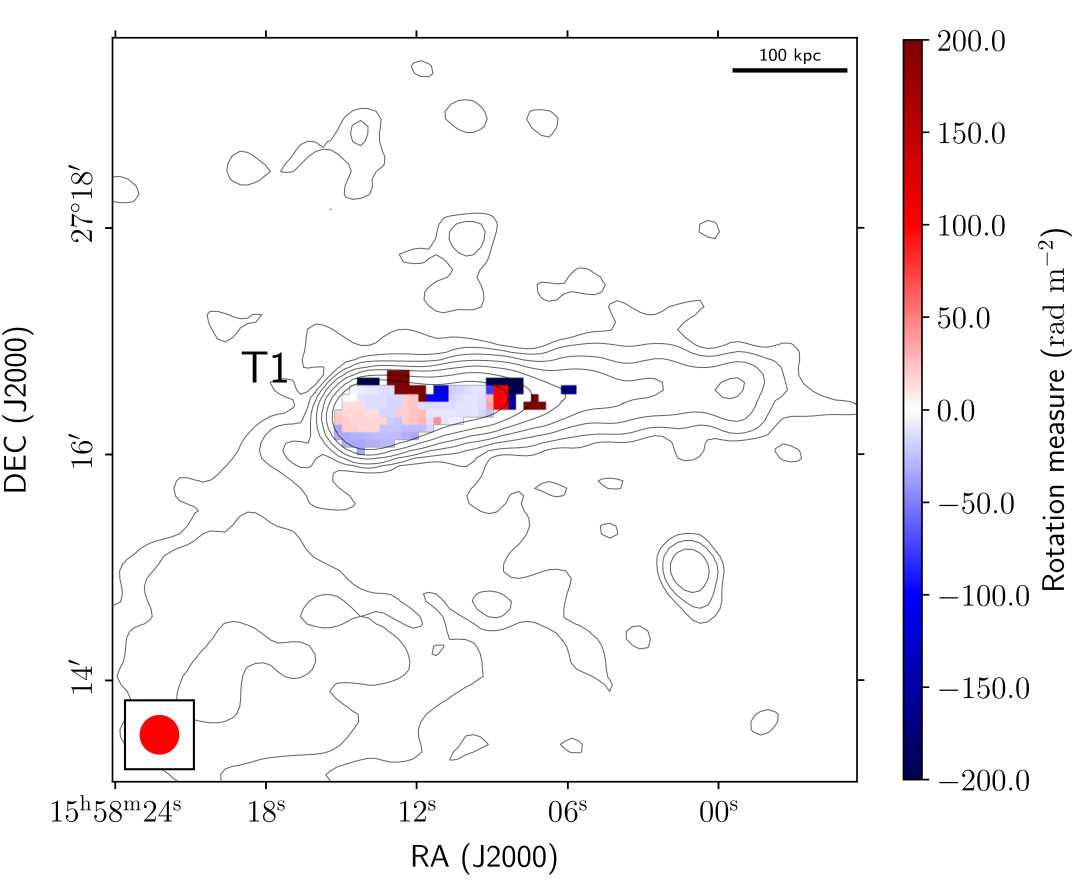}
        \caption{}
        \label{subfig:RM-T1}
    \end{subfigure}
    \hspace{-0.8cm}
    \begin{subfigure}{0.52\linewidth}
        \centering
        \includegraphics[width=0.91\linewidth]{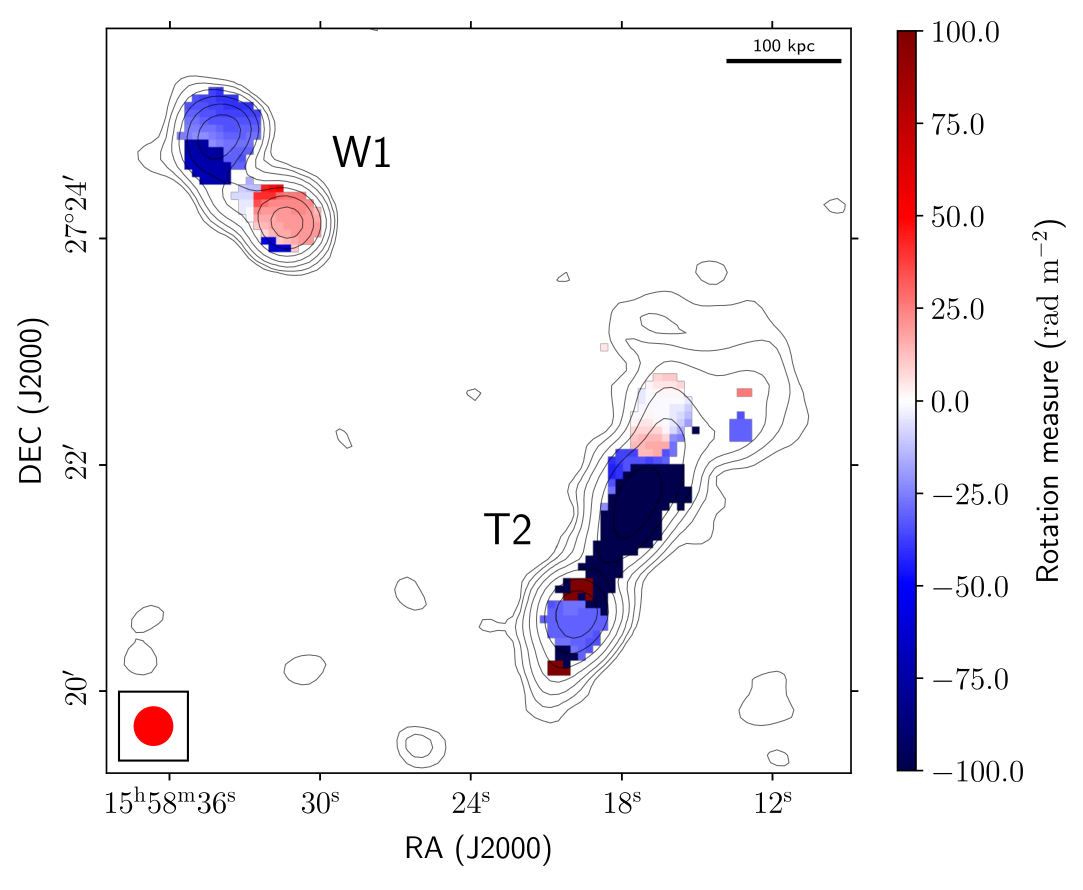}
        \caption{}
        \label{subfig:RM-T2-W1}
    \end{subfigure}
    \begin{subfigure}{0.52\linewidth}
        \hspace{-0.2cm}
        \includegraphics[width=0.91\linewidth]{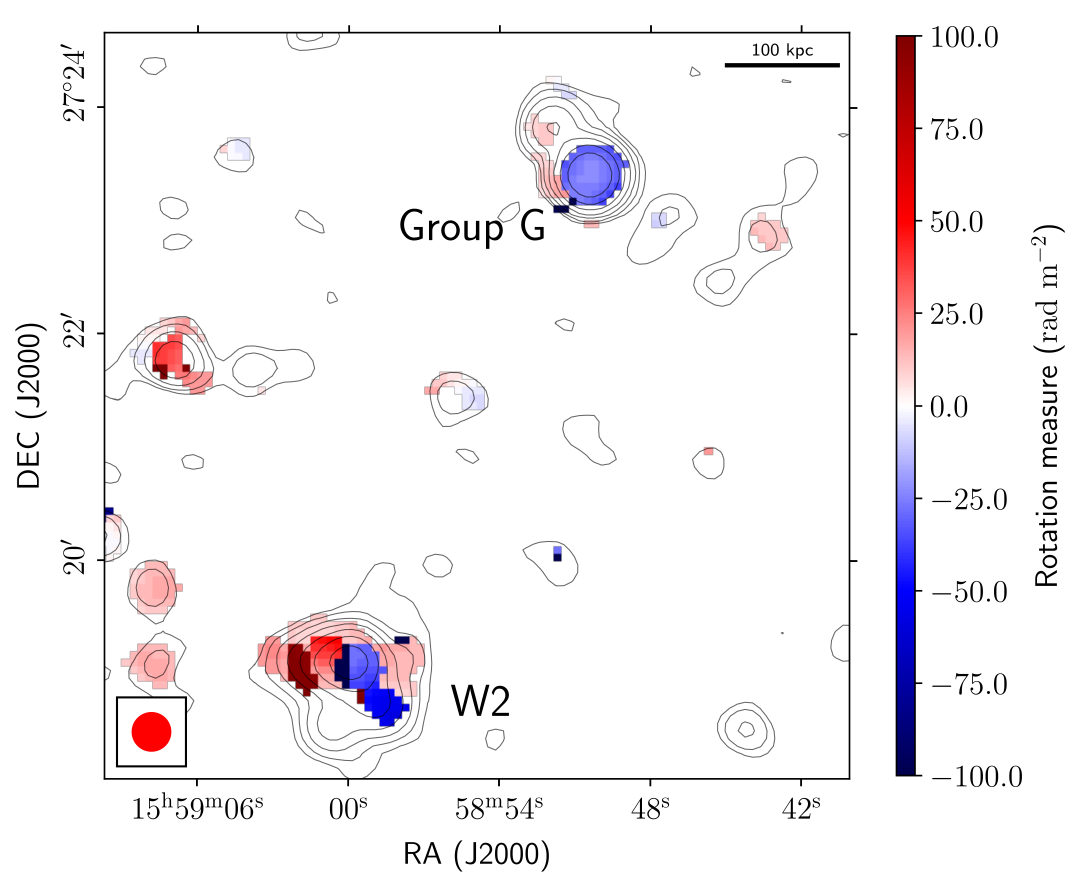} 
        \caption{}
        \label{subfig:RM-W2-GG}
    \end{subfigure}
    \hspace{-0.8cm}
    \begin{subfigure}{0.52\linewidth}
        \centering
        \includegraphics[width=0.91\linewidth]{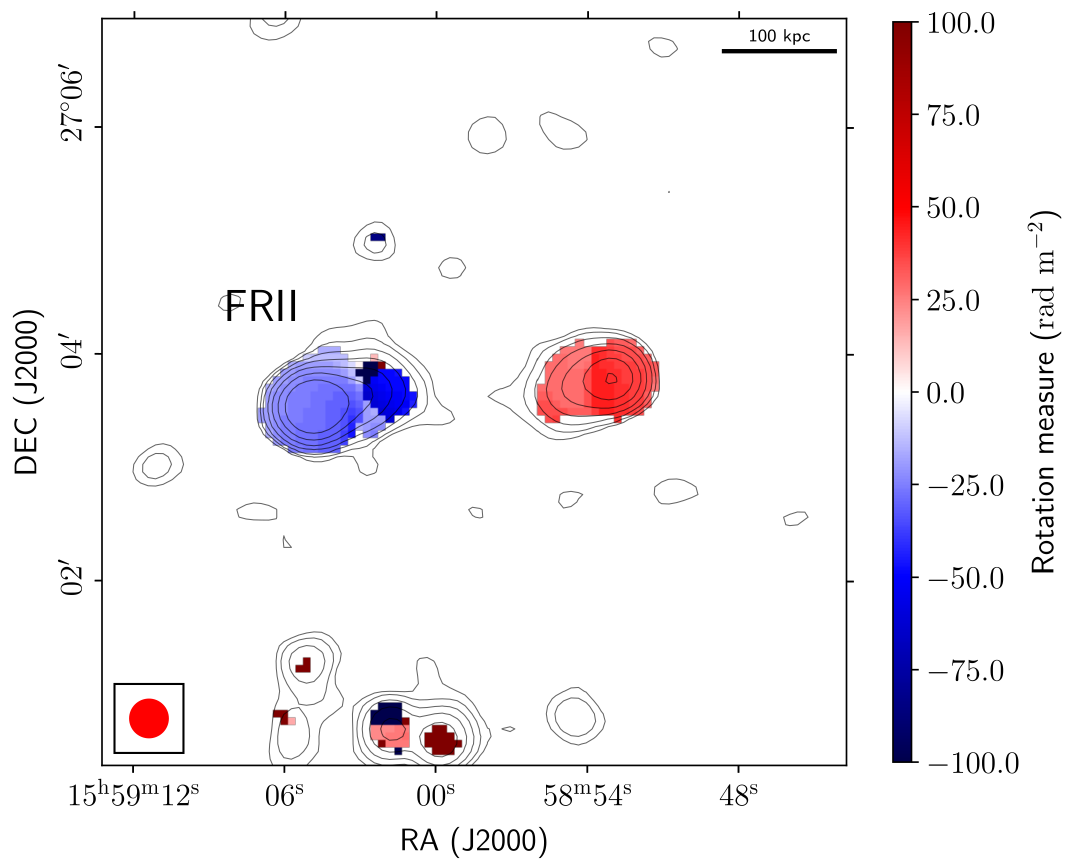} 
        \caption{}
        \label{subfig:RM-FRII}
    \end{subfigure}
    \caption{Zoom of the RM map in correspondence of some sources analysed in this work with overlaid radio contours (1283 MHz), from $3\sigma_{\mathrm{I}}$ and scaling by a factor of 2. The $6\sigma_{\mathrm{QU}}$ and $3\sigma_{\mathrm{I}}$ detection thresholds were imposed in polarisation and in total intensity and only pixels above them are shown. Values were corrected for the Galactic foreground rotation. Panel (a): T1. Panel (b): T2 and W1. Panel (c): W2 and Group G. Panel (d): FRII. Positive values refer to the MF orientation along the LOS towards the observer, whereas negative values refer to an orientation away from the observer. Spatial scales and resolution beam are reported on the edges of the images.}
    \label{fig:RM-sources}
\end{figure*}

Following further inspection, we noticed a bubble-like structure seen in projection in the top-left side of the map, extending beyond the cluster $\rm R_{500}$ and hereafter referred to as a ‘Galactic bubble’. The polarised radiation has no evidence of a diffuse Stokes I counterpart and was detected at a $6\sigma_{\mathrm{QU}}$ significance. Given its distance from the pointing centre, we first investigated the possibility that this feature originates from polarisation leakage. We refer to \citealt{Hugo_2024}\footnote{available at \url{https://doi.org/10.48479/bqk7-aw53}.} for additional details on the issue of the off-axis leakage in MeerKAT L-band observations. In Fig. \ref{fig:max-PI}, we show the 0.23${\degree}$ ($\sim 1381$ kpc at the cluster redshift) separation from the pointing centre, beyond which the polarisation leakage starts to be important at frequencies above 1.4 GHz.  We checked the presence of polarised emission in the region of the Galactic bubble as a function of frequency, using the Stokes Q and U cubes. This feature shows up at frequencies below 1.1 GHz, which are less affected by polarisation leakage, therefore, the instrumental effect is not the primary cause of such emission. We conclude that this feature could be associated with a Galactic component, even though it is at high Galactic latitude ($48.7\degree)$. To the best of our knowledge, there is no information about this Galactic structure in the literature. However, polarised Galactic emissions without total intensity counterparts have been previously reported by radio observations at low frequencies \citep{Balboni_2023}. \\
\begin{figure*}
    \begin{subfigure}{0.52\linewidth}
        \hspace{-0.2cm}
        \includegraphics[width=0.91\linewidth]{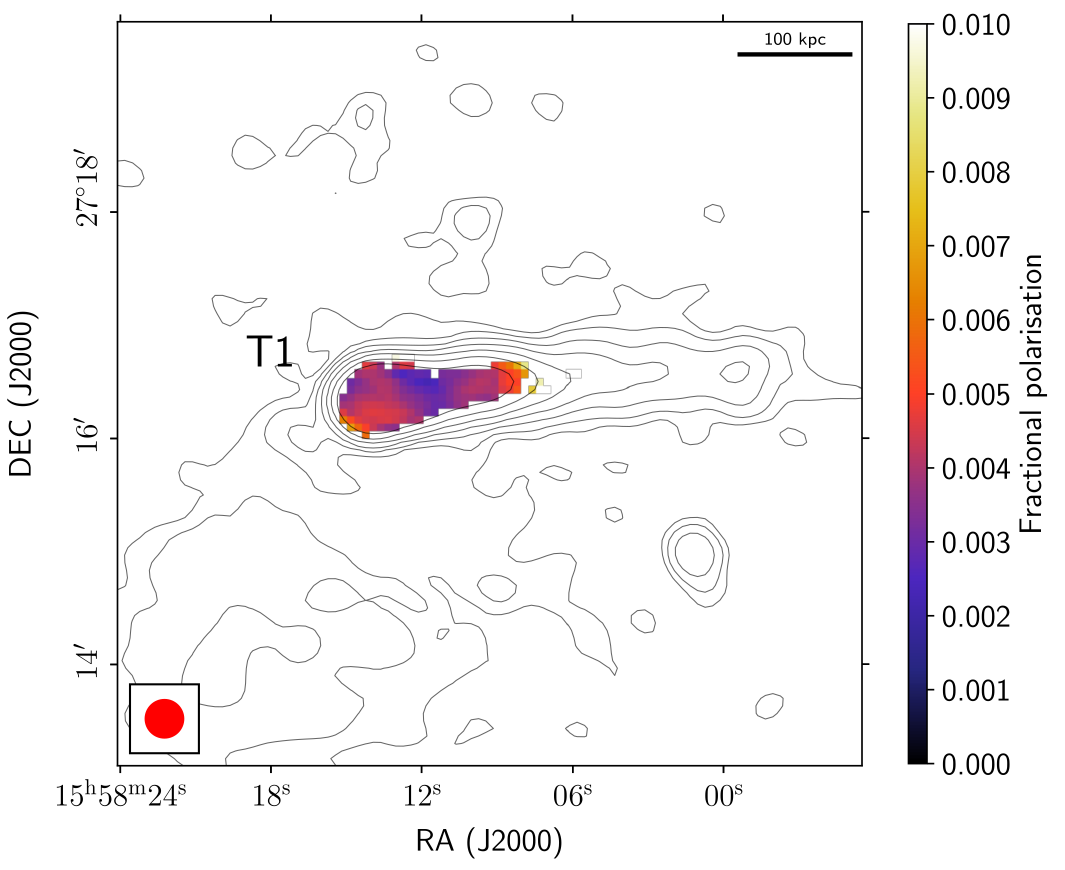}
        \caption{}
        \label{subfig:Fp-T1}
    \end{subfigure}
    \hspace{-0.8cm}
    \begin{subfigure}{0.52\linewidth}
        \vspace{-1cm}
        \centering
        \includegraphics[width=0.91\linewidth]{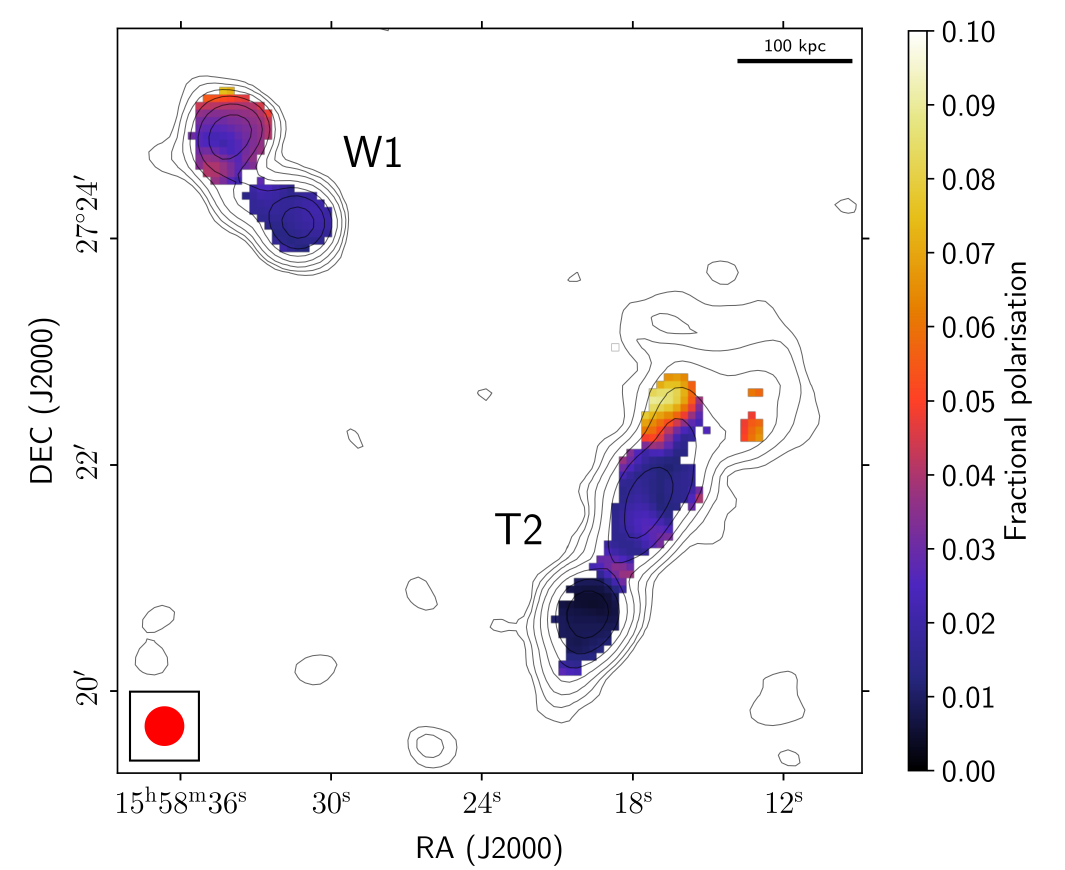}
        \caption{}
        \label{subfig:Fp-T2-W1}
    \end{subfigure}
    \begin{subfigure}{0.52\linewidth}
        \hspace{-0.2cm}
        \includegraphics[width=0.91\linewidth]{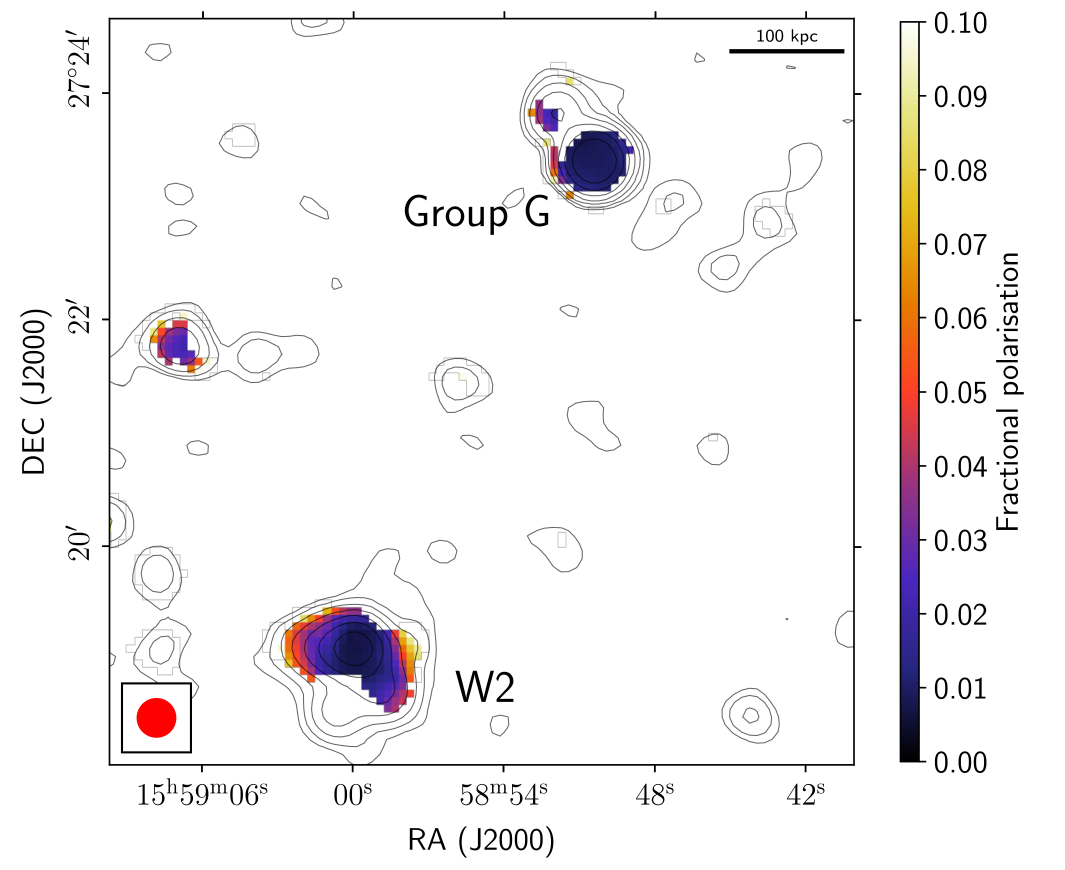} 
        \caption{}
        \label{subfig:Fp-W2-GG}
    \end{subfigure}
    \hspace{-0.8cm}
    \begin{subfigure}{0.52\linewidth}
        \centering
        \includegraphics[width=0.91\linewidth]{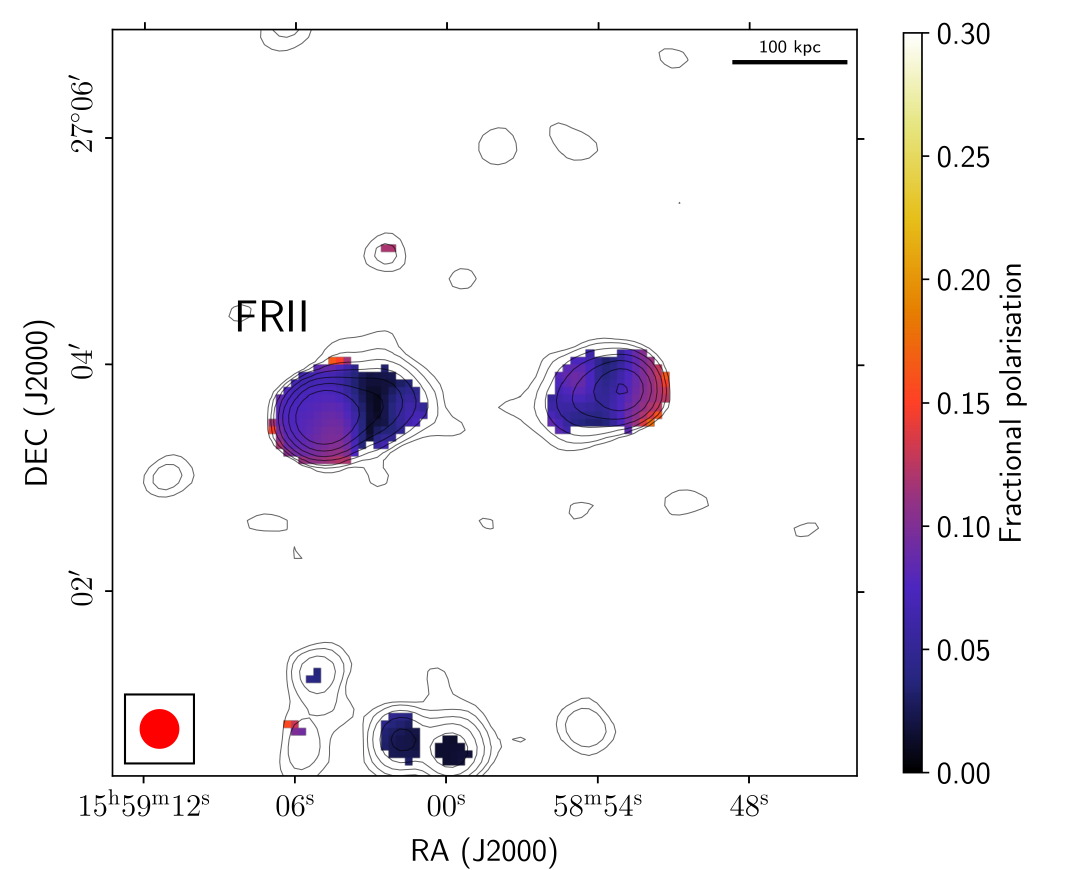} 
        \caption{}
        \label{subfig:Fp-FRII}
    \end{subfigure}
    \caption{Zoom of the $\mathrm{F_p}$ map in correspondence of some sources analysed in this work with overlaid radio contours (1283 MHz), from $3\sigma_{\mathrm{I}}$ and scaling by a factor of 2. The $6\sigma_{\mathrm{QU}}$ and $3\sigma_{\mathrm{I}}$ detection thresholds were imposed in polarisation and in total intensity and only pixels above them are shown. Values were corrected for the Ricean bias. Panel (a): T1. Panel (b): T2 and W1. Panel (c): W2 and Group G. Panel (d): FRII. Spatial scales and resolution beam are reported on the edges of the images.}
    \label{fig:Fp-sources}
\end{figure*}
\\
To investigate the possible presence of complex FDF indicating multiple Faraday components along the LOS, we  selected some radio galaxies in the field and extracted at the location of the brightest polarised pixel the dirty FDF spectrum, the cleaned FDF spectrum after deconvolution, and the clean components. 
In case of Faraday-simple objects, $|\tilde{F}(\phi)|$ peaks at the $\phi$ corresponding to the RM of the source and if more than one is seen through the same LOS, each peak corresponds to a source. Fig. \ref{subfig:FDF-T1} displays the case of the central radio galaxy, T1, and in polarisation a clear double peak and other minor peaks are visible, at a higher $|\phi|$, above the $8\sigma_{\mathrm{QU}}$ threshold. 
We note that the algorithm has identified multiple components around $0 \ \mathrm{rad \ m^{-2}}$, whose superposition gives rise to the dirty FDF primary lobe. In this case, the FWHM of the RMTF, $45.87 \ \mathrm{rad \ m^{-2}}$, is larger with respect to separation of the two peaks in the clean spectrum; that is, we were not able to resolve the ‘Faraday-complex' spectrum. In this case, the polarised emission is spread at different $\phi$ and the values of $|\tilde{F}(\phi_{\rm peak})|$ and of $\phi_{\rm peak}$ are not sufficient to describe the polarisation and the rotation effect experienced by the radiation. One possible scenario that could explain these results is the orientation of the tail of the radio galaxy seen in projection: the bent jets and lobes could be superimposed on a single LOS and consist of a combination of Faraday-rotating and synchrotron-emitting media. Similar cases for other radio galaxies in the field are reported in App. \ref{AppA}. Figure \ref{subfig:FDF-Sbkg}, instead, refers to a source in the background and is an evident example of a Faraday-simple spectrum.\\
\begin{table*}[!htb]
    \vspace{0.3cm}
    \centering
    \caption{RM and polarisation properties of the known sources detected in polarisation.}
    \label{tab:radvalues}
    \begin{tabular}{c c c c c c c c c c c c c c c c}
        \hline
        \noalign{\smallskip}
        \noalign{\smallskip}
         Source & & & Distance & & & $\mathrm{n_{beam}}$ & & & $\langle \mathrm{RM} \rangle \pm \mathrm{err_{\langle RM \rangle}}$ & & & $\mathrm{\sigma_{RM}} \pm \mathrm{err_{\sigma_{RM}}}$ & & & $\mathrm{F_p} \pm \mathrm{err_{F_p}}$ \\
          & & & [kpc] & & & & & &$\mathrm{[rad \ m^{-2}]}$ & & & $\mathrm{[rad \ m^{-2}]}$ & & & $[\%]$ \\
         (1) & & & (2) & & & (3) & & & (4) & & & (5) & & & (6)\\
         \noalign{\smallskip}
         \hline
         \noalign{\smallskip}
         \noalign{\smallskip}
         T1 & & & 279.31 & & & 5 & & & -16.6 $\pm$ 80.4 & & & 189.9 $\pm$ 56.8 & & & 0.45 $\pm$ 0.08 \\ 
         \noalign{\smallskip}
         T2 & & & 600.77 & & & 10 & & & -84.3 $\pm$ 39.4 & & & 127.8 $\pm$ 28.4 & & & 2.62 $\pm$ 0.21 \\ 
         \noalign{\smallskip}
         W1 & & & 1046.68 & & & 6 & & & -19.7 $\pm$ 13.9 & & & 34.7 $\pm$ 9.8 & & & 2.76 $\pm$ 0.36 \\ 
         \noalign{\smallskip}
         W2 & & & 989.65 & & & 6 & & & 9.8 $\pm$ 43.3 & & & 110.1 $\pm$ 30.6 & & & 5.13 $\pm$ 0.87 \\ 
         \noalign{\smallskip}
         Group G & & & 1150.19 & & & 3 & & & -16.6 $\pm$ 12.8 & & & 23.6 $\pm$ 9.1 & & & 5.15 $\pm$ 1.75 \\
         \noalign{\smallskip}
         FR II & & & 1437.59 & & & 11 & & & -12.1 $\pm$ 25.7 & & & 88.9 $\pm$ 18.2 & & & 7.14 $\pm$ 1.14\\ 
         \noalign{\smallskip}
         \hline
    \end{tabular}
    \tablefoot{(1): identification name of the source as shown in Fig. \ref{subfig:Stokes-I_zoom}; (2): projected distance of the source from the cluster centre; (3): number of beams of the source; (4): average RM of the source with associated error; (5): RM dispersion of the source with associated error; (6): fractional polarisation of the source with associated error.}
\end{table*}
\begin{table*}[!htb]
    \vspace{0.3cm}
    \centering
    \caption{RM and polarisation properties of the unknown sources detected in polarisation.}
    \label{tab:ukradvalues}
    \begin{tabular}{c c c c c c c c c c c c cc c c c c}
        \hline
        \noalign{\smallskip}
        \noalign{\smallskip}
         Source & & & Distance & & & $\mathrm{n_{beam}}$ & & & $\langle \mathrm{RM} \rangle \pm \mathrm{err_{\langle RM \rangle}}$ & & & $\mathrm{\sigma_{RM}} \pm \mathrm{err_{\sigma_{RM}}}$ & & & $\mathrm{F_p} \pm \mathrm{err_{F_p}}$\\
          & & & [kpc] & & & & & & $\mathrm{[rad \ m^{-2}]}$ & & & $\mathrm{[rad \ m^{-2}]}$ & & & $[\%]$\\
         (1) & & & (2) & & & (3) & & & (4) & & & (5) & & & (6)\\
         \noalign{\smallskip}
         \hline
         \noalign{\smallskip}
         \noalign{\smallskip}
         BL1 & & & 1503.94 & & & 7 & & & -15.1 $\pm$ 7.8 & & & 21.3 $\pm$ 5.5 & & & 4.53 $\pm$ 0.73\\ 
         \noalign{\smallskip}
         TR1 & & & 1789.07 & & & 5 & & & 6.6 $\pm$ 20.1 & & & 45.2 $\pm$ 14.2 & & & 3.68 $\pm$ 1.18\\ 
         \noalign{\smallskip}
         BR1 & & & 1760.69 & & & 3 & & & 16.3 $\pm$ 2.4 & & & 4.1 $\pm$ 1.7 & & & 5.41 $\pm$ 1.95\\ 
         \noalign{\smallskip}
         BR2 & & & 2067.57 & & & 7 & & & 20.8 $\pm$ 3.1 & & & 8.2 $\pm$ 2.2 & & & 6.00 $\pm$ 1.02 \\ 
         \noalign{\smallskip}
         TL1 & & & 1699.23 & & & 11 & & & -2.0 $\pm$ 16.5 & & & 56.3 $\pm$ 11.7 & & & 8.22 $\pm$ 1.40 \\ 
         \noalign{\smallskip}
         TL2 & & & 1665.54 & & & 7 & & & 18.6 $\pm$ 9.7 & & & 26.9 $\pm$ 6.9 & & & 8.60 $\pm$ 1.55\\ 
         \noalign{\smallskip}
         TL3 & & & 1643.40 & & & 9 & & & 6.8 $\pm$ 6.5 & & & 19.6 $\pm$ 4.6 & & & 7.66 $\pm$ 1.30 \\ 
         \noalign{\smallskip}
         TL4 & & & 1600.49 & & & 5 & & & 8.0 $\pm$ 17.7 & & & 40.4 $\pm$ 12.5 & & & 8.37 $\pm$ 2.09 \\ 
         \noalign{\smallskip}
         \hline
    \end{tabular}
    \tablefoot{(1): identification name associated with one of the four quadrants of the image in which the sources are located (BL: bottom-left; TR: top-right; BR: bottom-right; TL: top-left); (2), (3), (4), (5), (6) columns description follows Table \ref{tab:radvalues}.}
\end{table*}
\\
In the total RM map, we observed values ranging from $-250$ to $+250 \ \mathrm{rad \ m^{-2}}$ (smaller with respect to the values at which we are sensitive; see Table \ref{tab:RMSparameters}) according to the different intensity and orientation of the MF along the LOS and a patchy appearance, especially in correspondence to the embedded resolved radio galaxies, due to the MF components fluctuating on scales smaller than the source size. The RM distribution of the external resolved sources, instead, looks more uniform, which indicates that the MF substantially changes on scales larger than the source extension. Zooms of the RM map are provided in Fig. \ref{fig:RM-sources} and Fig. \ref{fig:RM-sources-2}, in correspondence to the sources considered in the further analysis (see Sect. \ref{Radial profiles}).\\
\\
In the total $\mathrm{F_p}$ map, we observe a clear distribution of the fractional polarisation, with values ranging from 0.25\% to 30\% and increasing towards the edges of the cluster. The radio galaxies seen in projection in the central regions (e.g. T1 and T2) show the lowest values and this is in agreement with larger RM dispersion in the cluster center due to a more intense MF. The highest percentage values are concentrated in the region of the Galactic bubble. It is important to notice that in the case of T1, only one of the two primary peaks at low $|\rm RM|$ shown in Fig. \ref{subfig:FDF-T1} was taken into account for the creation of the polarisation map. Therefore, only a fraction of the polarised emission was considered and the true value of $\mathrm{F_p}$ is likely underestimated. Zooms of the fractional polarisation map are provided in Fig. \ref{fig:Fp-sources} and Fig. \ref{fig:Fp-sources-2} in correspondence to the same above selected sources.

\subsection{Radial profiles} \label{Radial profiles}
\begin{figure*}[!htb]
    \centering
    \includegraphics[width=0.7\linewidth]{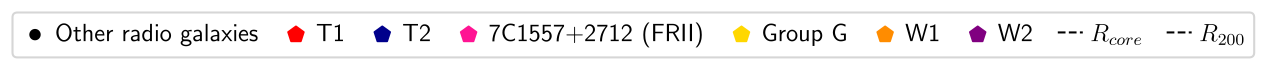}
    \vspace{-0.3cm}
\end{figure*}
\begin{figure*}
    \begin{minipage}[b]{0.36\textwidth}
    \hspace{-0.12\textwidth}
        \begin{subfigure}[b]{\linewidth}
            \includegraphics[width=\linewidth]{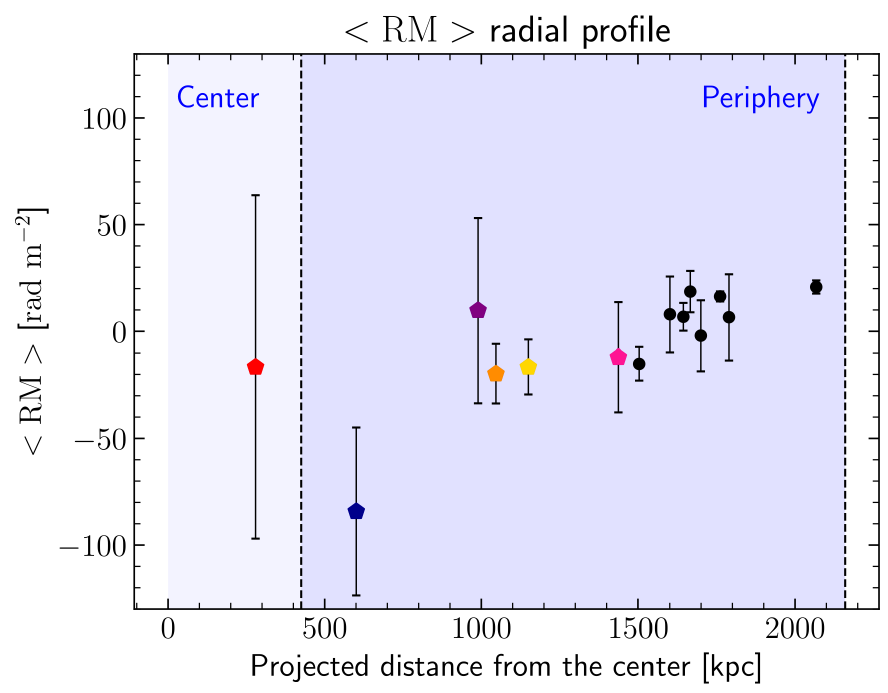}
            \caption{}
            \label{subfig:averageRM_pjdist}
        \end{subfigure}
    \end{minipage}
    \begin{minipage}[b]{0.3525\textwidth}
    \hspace{-0.1\textwidth}
        \begin{subfigure}[b]{\linewidth}
            \includegraphics[width=\linewidth]{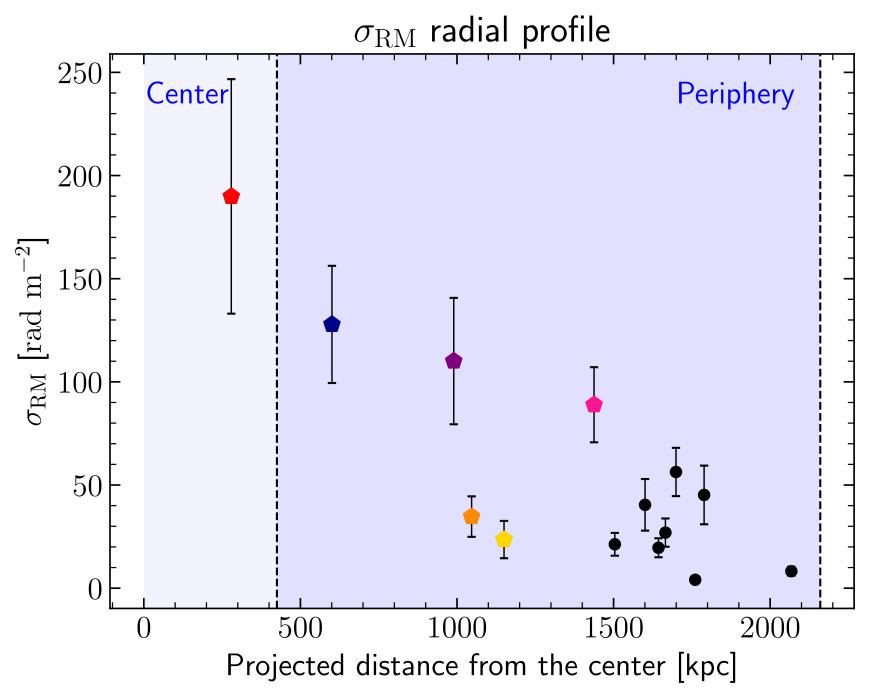}
            \caption{}
            \label{subfig:RMdisp_pjdist}
        \end{subfigure}
    \end{minipage}
    \begin{minipage}[b]{0.355\textwidth}
    \hspace{-0.1\textwidth}
        \begin{subfigure}[b]{\linewidth}
            \includegraphics[width=\linewidth]{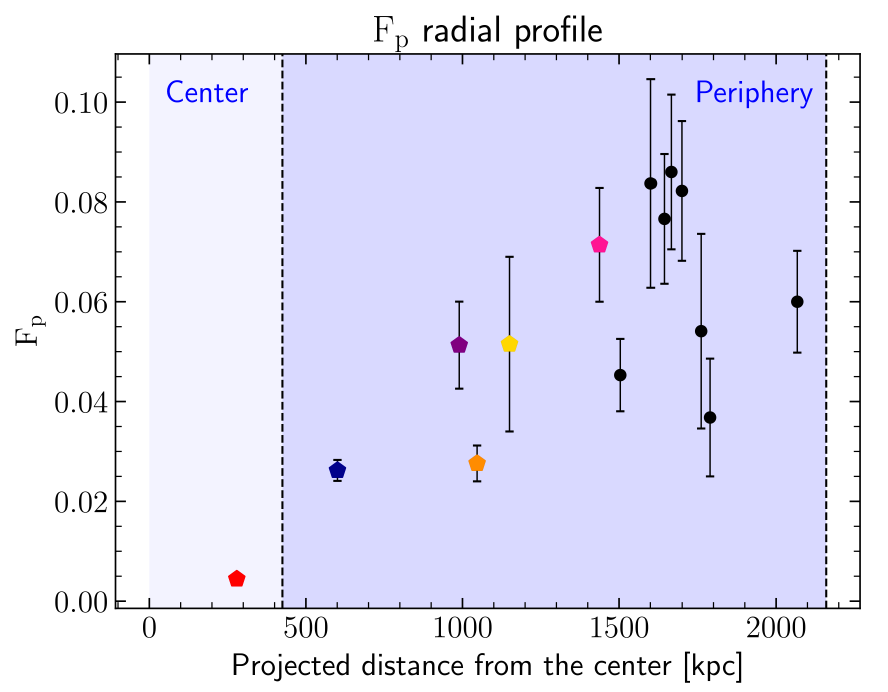}
            \caption{}
            \label{subfig:Fp_pjdist}
        \end{subfigure}
    \end{minipage}
    \caption{$\langle \mathrm{RM} \rangle$, $\sigma_{\mathrm{RM}}$, and $\mathrm{F_p}$ as a function of the projected distance from the cluster centre. The known sources are represented with the coloured pentagons, whereas the other radio galaxies (listed in Table \ref{tab:ukradvalues}) with the black dots. Panel (a): $\mathrm{\langle RM \rangle}$ radial profile. Panel (b): $\sigma_{\mathrm{RM}}$ radial profile. Panel (c): ${\mathrm{F_p}}$ radial profile. The errors are computed following Eqs. \ref{eq:RMerr} and \ref{eq:Fperr}. Reference distances, such as $\mathrm{R_{\rm core}=425.60 \pm 20.03 \ kpc}$ \citep{Henry_1996} and $\mathrm{R_{200}=2.16 \pm 0.08 \ Mpc}$ \citep{Munari_2014}, are reported with the dashed lines to discriminate the central from the peripheral regions of the cluster.}
    \label{fig:Radialprofiles}
\end{figure*}
We studied the average RM, $\langle \mathrm{RM} \rangle$, RM dispersion, $\sigma_{\mathrm{RM}}$, and fractional polarisation, $\mathrm{F_p}$, variation as a function of the projected distance from the cluster centre. In Tables \ref{tab:radvalues} and \ref{tab:ukradvalues}, we summarise the properties of the sources for this analysis. The known sources are labeled as in Fig. \ref{subfig:Stokes-I_zoom}, while for the unknown ones we chose an identification name according to the quadrant of the image in which they are located (e.g. bottom-left:\ BL; top-right:\ TR; bottom-right:\ BR; and top-left:\ TL). These radio galaxies are located beyond the off-axis polarisation leakage threshold (see Fig. \ref{fig:max-PI}), therefore, the associated polarised emission could be overestimated. For this reason, these sources are used for the global considerations on the radial profiles, but not taken into account for deriving the MF constraints (Sects. \ref{MF scale} and \ref{MF peak}). The radial distance of each source is computed as the projected distance between the X-ray peak and the brightest polarised pixel detected at the source position. The errors in $\langle \mathrm{RM} \rangle$ and in $\sigma_{\mathrm{RM}}$ are calculated as
\begin{equation}
    \mathrm{err_{\langle RM \rangle}}=\dfrac{\sigma_{\mathrm{RM}}}{\sqrt{\rm n_{\mathrm{beam}}}}, \ \ \mathrm{err_{\sigma_{RM}}}=\dfrac{\sigma_{\mathrm{RM}}}{\sqrt{2 \rm n_{\mathrm{beam}}}},
    \label{eq:RMerr}
\end{equation}
where $\rm n_{\mathrm{beam}}$ refers to the number of independent beams over which the RM is calculated. Only sources with $\rm n_{\mathrm{beam}}\geq 3$ have been considered statistically significant for the analysis. The errors in $\mathrm{F_p}$ are defined based on the propagation of the errors as
\begin{equation}
    \mathrm{err_{F_p}}=\mathrm{F_p}\sqrt{\left(\dfrac{\sigma_{\mathrm{QU}}}{\mathrm{P}}\right)^2+\left(\dfrac{\sigma_{\mathrm{I}}}{\mathrm{I}}\right)^2}.
    \label{eq:Fperr}
\end{equation}
Equation \ref{eq:RMerr} represents the scatter of $\langle \mathrm{RM} \rangle$ and $\sigma_{\mathrm{RM}}$ respectively, while Eq. \ref{eq:Fperr} refers to the uncertainty in $\mathrm{F_p}$ due to the S/N ratio.

In Fig. \ref{subfig:averageRM_pjdist}, we show the radial profile of $\langle \mathrm{RM} \rangle$ for both the cluster radio sources and the radio galaxies in the background. The modulus of the average RM, $|\langle \mathrm{RM} \rangle|$, decreases as a function of the projected distance from the centre, suggesting that the MF becomes less intense towards larger distances. Only T1 presents lower values with respect to the expected trend, and this could be due to projection effects (as the radio galaxy may be located in the peripheral regions of the cluster, but seen in projection in the core) or to the averaging of positive and negative RM contributions. As discussed in Sect. \ref{RM Synthesis Results}, T1 shows two peaks around $\phi = 0 \ \mathrm{rad \ m^{-2}}$ in the clean Faraday spectrum (see Fig. \ref{subfig:FDF-T1}). 
However, due to limitations in instrumental sensitivity, we are unable to resolve both components and only one of the two peaks is considered when creating the RM map. 

Since the RM can assume both positive and negative values, according to the orientation of the MF along the LOS, the $\langle \mathrm{RM} \rangle$ can be zero even in the presence of a strong MF. Consequently, the dispersion of the RM, $\sigma_{\mathrm{RM}}$, is a better proxy of the MF strength and will be used later on during the comparison with the simulations. We show in Fig. \ref{subfig:RMdisp_pjdist}, the radial trend of $\sigma_{\mathrm{RM}}$, going from 100s $\mathrm{rad \ m^{-2}}$ in the central regions and declining to 10s $\mathrm{rad \ m^{-2}}$ in the periphery. As this quantity is much more sensitive to the variation of the MF, this profile shows the MF decrease with distance deduced from Fig. \ref{subfig:averageRM_pjdist} more clearly.

Finally, the fractional polarisation radial profile (Fig. \ref{subfig:Fp_pjdist}) confirms that radio galaxies seen in projection near the cluster centre suffer higher depolarisation than peripheral ones, where the MF is expected to become less intense and producing lower values of the RM dispersion (e.g. \citealt{Osinga_2022}). 
In addition, T1 is showing very low values of $\mathrm{F_p}$  
as a direct consequence of a multiple intervening material, which is spreading the polarised emission at different $\phi$ (see Fig. \ref{subfig:FDF-T1}). In this case, only a part of the total signal is considered during the production of the $\mathrm{F_p}$ map.

We note that we plotted radial profiles, thereby assuming a spherical symmetry. While we acknowledge that the X-ray surface brightness distribution is not spherical (Fig. \ref{subfig:X-ray}), implying a more complex gas density distribution, the main finding here is that $\sigma_{\mathrm{RM}}$ clearly decreases at larger distances from the cluster centre. In the next section, we aim to address the $\sigma_{\mathrm{RM}}$  dependence on thermal gas density and integration length more carefully.

\section{Simulations and magnetic field profile} \label{Simulations}
To recover the MF intensity along the LOS, in principle, we need to invert Eq. \ref{eq:RM} -- once we know the extent of the Faraday-rotating medium along the LOS and the thermal electron gas density distribution from X-ray observations. It is not trivial to perform such an operation and the main uncertainty is related to the 3D structure of the MF. We used semi-analytical simulations of 3D MF structure, collapsed into 2D, to generate synthetic and mock RM maps by solving Eq. \ref{eq:RM}. These are then compared statistically with the observational results in order to reconstruct the MF profile responsible for the observed values. This method has been used so far to constrain the MF in the ICM of both relaxed and merging (or NCC) clusters (\citealt{Murgia_2004}; \citealt{Govoni_2006}; \citealt{Guidetti_2008}; \citealt{Bonafede_2010}, \citeyear{Bonafede_2013}; \citealt{Vacca_2012}; \citealt{Govoni_2017}), but also of radio relics (e.g. \citealt{Bonafede_2013}; \citealt{Stuardi_2021}; \citealt{DeRubeis_2024} from polarisation studies).\\ 
\\
In this section, we  describe the simulations used in order to constrain the MF properties of A2142. The adopted 3D MF model and the combinations of parameters used to build the simulations are explained in Sect. \ref{MIROcode}. In Sect. \ref{MF scale}, we compare the values extracted from the mock RM maps with the observations adopting a statistical approach. In Sect. \ref{MF peak}, we explore other MF models, determine the best-fit combination of parameters, and constrain the cluster MF intensity profile.

\subsection{Magnetic field modelling: MIRO' code} \label{MIROcode}
Both radio observations (e.g. \citealt{Murgia_2004}; \citealt{Bonafede_2010}) and cosmological simulations (e.g. \citealt{Vazza_2018}; \citealt{DF_2019}) have suggested that the MF changes on a range of physical scales. To accomplish this, we used the modified version of the \texttt{MIRO'} code (\citealt{Bonafede_2013}; \citealt{Stuardi_2021}), on A2142. Briefly, the code is designed to start by creating 3D MF and gas density models and to produce 2D RM maps according to Eq. \ref{eq:RM}.\\
\\
We proceed with a more detailed description of the algorithm step by step below. 
\begin{itemize}
    \setlength{\itemsep}{-0.5pt}
    \item[-] First, it creates a 3D gas density model based on the universal electron density profile recently determined by \citealt{Ghirardini_2019}. The authors  exploited X-ray observations of 12 XCOP galaxy clusters (including A2142) and fit their de-projected density profiles with the functional form from \citealt{Vikhlinin_2006} to parameterise the radial behavior as      \begin{equation}
        \mathrm{n_e}^2(x)=\mathrm{n_0}^2\dfrac{(x/r_c)^{-\alpha}}{(1+x^2/r_c^2)^{3\beta-\alpha/2}}\dfrac{1}{(1+x^{\gamma}/r_s^{\gamma})^{\epsilon/\gamma}},
        \label{eq:densityprofile}
    \end{equation}
    where $x = \rm R/R_{500}$ and $\gamma = 3$. In this way, the code takes in input $\rm R_{500}$ and the dynamical state of the cluster (for A2142: NCC) to set the best-fit parameters of the profile. The specific values for NCC clusters are $\log(\rm n_0)=-4.9$, $\log(r_c)=-2.7$, $\alpha=0.70$, $\beta=0.39$, $\log(r_s)=-0.51,$ and $\epsilon=2.60$ ($\rm n_0$ has a dimension of $10^{-3}\mathrm{cm}^{-3}$, whereas the others are dimensionless parameters). This model is able to reproduce the observed radial steepening from the core out to two decades in radius (2$\rm R_{500}$),
    \item[-] Secondly, it generates a 3D MF model from the analytical power spectrum derived by \citealt{DF_2019}, which represents a much more realistic MF energy distribution expected in galaxy clusters with respect to the Kolmogorov power law spectrum, $\rm \propto k^{-n}$ (e.g. \citealt{Murgia_2004}; \citealt{Bonafede_2010}; for a comparison between the two spectra, see \citealt{Stuardi_2021} and \citealt{DeRubeis_2024}). Using cosmological MHD simulations, \citealt{DF_2019} found that the 1D magnetic spectra of seven simulated galaxy clusters can be well fitted to 
    \begin{equation}
        \rm E_B(k)\propto k^{3/2}\Biggr[1-\mathrm{erf}\Biggr(B \ \mathrm{ln}\dfrac{k}{C}\Biggr)\Biggr],
        \label{eq:PowerSpectrum}
    \end{equation}
    where $\rm k=\sqrt{\sum_ik_i^2}$ (with $\rm i=1,2,3$) is the wavenumber corresponding to the physical scale of the MF fluctuations (e.g. $\rm \Lambda \propto 1/k$), B is a parameter related to the width of the spectrum and C is the wavenumber corresponding to the peak of the spectrum. Both B and C depend on the dynamical state of the cluster, are the best-fit values derived from the simulated clusters analysed in \citealt{DF_2019}  and are taken in input by the code.\par
    The MF generated by the code is by definition divergence-free, with Gaussian components, $B_i$, having $\langle B_i \rangle = 0$ and $\sigma^2_{B_i}=\langle B_i^2\rangle$.
    The radial profile of the magnitude of the MF is expected to scale with the thermal electron density as 
    \begin{equation}
        \rm |\mathbf{B}(r)| \propto n_e(r)^{\eta}, 
        \label{eq:etaB}
    \end{equation}
    where $\eta$ is assumed to be equal to 0.5 \citep{Bonafede_2010}.  
    The normalisation of the MF distribution is finally obtained imposing that the MF averaged over the cluster volume is a scalar value, $\rm B_{\rm norm}$, and will be determined during the comparison with the observations;
    \item[-] Finally, the code produces a 2D RM map, integrating numerically the obtained thermal electron density and MF profiles along the $z$ axis of the cube, starting from the centre of the cluster.
\end{itemize}
Overall, the MF model depends on six parameters: the radial slope, $\eta$, the minimum and maximum spatial scales of the fluctuations, $\Lambda_{\rm min}$ and $\Lambda_{\rm max}$, the normalisation, $\rm B_{\rm norm}$, and the B and C parameters of the MF power spectrum.\\
\\
The simulation takes as input both the size of the simulated box and the cell resolution.  To match the observations, we  chose a pixel scale of 3.4 kpc (i.e. 2$^{\prime\prime}$ at the cluster redshift) and to produce a $829^3$ pixels cube to reach a cluster dimension of $\sim 2.8^3 \ \mathrm{Mpc}^3$ (i.e. out to $\rm R_{500}$ from the cluster centre). The maximum fluctuating scale of the MF components in Fourier space, $\rm k_{\rm max}$, is defined as $\rm k_{\rm max} =(829\cdot0.5)-1\sim413$. This gives 
\begin{equation}
    \Lambda_{\rm min}=\dfrac{829\cdot3.4}{\rm k_{\rm max}} \sim 7 \ \mathrm{kpc}.
\end{equation}
Regarding the power spectrum, B and C are respectively assumed to be 1.1 and 5.0 $\mathrm{Mpc}^{-1}$. According to \citealt{DF_2019}, C is computed over a box of 2 Mpc per side, while in our simulations the cube extends up to 2.8 Mpc. This translates into a power spectrum that is similar to the one found in \citealt{DF_2019}, but that peaks at $\sim 280$ kpc. The choice of these two values is justified by the observations in the optical, radio, and X-ray band discussed in Sect. \ref{Abell2142}. The asymmetric gas distribution, the presence of cold fronts, and substructures revealed by weak-lensing mass studies suggest that A2142 is not currently undergoing a major merger, but it is seen at least $1-2$ Gyr after the first event and has possibly experienced several minor mergers. Thus, we assumed $z_{\rm last}=0.1$, which corresponds to a post-merger dynamical state (see ID E1 in Table 1 of \citealt{DF_2019}). We summarise the fixed parameters of the simulations in Table \ref{tab:Sim specifics}.\\
\begin{table}[!htb]
    \vspace{-0.3cm}
    \caption{List of fixed parameters adopted for the simulations.}
    \begin{center}
        \begin{tabular}{ c c c c c c c} 
            \hline
            \noalign{\smallskip}
            \noalign{\smallskip}
            Size of the box & & & 829 pxl & & &\\ 
            \noalign{\smallskip}
            Resolution & & & 3.4 kpc & & & \\
            \noalign{\smallskip}
            $\rm\Lambda_{min}$ & & & 7 kpc & & &\\ 
            \noalign{\smallskip}
            $\eta$ & & & 0.5 & & &\\
            \noalign{\smallskip}
            Dynamical state & & & NCC & & & (1) \\
            \noalign{\smallskip}
            $\rm R_{500}$ & & & 1409 kpc & & & (2) \\
            \noalign{\smallskip}
            B & & & 1.1 & & & (3) \\
            \noalign{\smallskip}
            C & & & 5.0 $\mathrm{Mpc}^{-1}$ & & &(3) \\
            \noalign{\smallskip}
            \hline
     \end{tabular}
     \tablefoot{(1): for the \citealt{Ghirardini_2019} density profile; (3) for the \citealt{DF_2019} MF power spectrum.}
     \tablebib{(2): \citealt{PC_2016}.}
     \label{tab:Sim specifics}
     \end{center}
     \vspace{-0.8cm}
\end{table}
\\
The final outputs of the simulations are: three cubes with the three components of the MF model ($\rm B_x$, $\rm B_y$, and $\rm B_z$), the gas density cube and a cube with the projected volume weighted mean density, the mean MF and RM computed along half the $z$ axis for each pixel, with size $\sim2.8^2 \ \mathrm{Mpc}^2$, and a resolution of 3.4 kpc.\\
\\
Our MF model considers a total of two free parameters, $\Lambda_{\rm max}$ and $\rm B_{\rm norm}$, that can be investigated through the comparison with our observations. We  chose the following sets of parameters, $\Lambda_{\rm max}=[50,470,940] \ \mathrm{kpc}$ and $\rm B_{\rm norm}=[0.7, 1.0, 1.5, 3.0] \ \mu \mathrm{G}$. For every combination, we  performed five simulations from different random seeds, since different realisations of the same model will correspond to different values of the RM at a given position\footnote{For one of the MF models, we ran ten different simulations. Upon examining the results, we found that the average radial profiles remained largely unchanged whether considering the full set or only half of the simulations. The MF strength values also remained within the statistical scatter. Therefore, to reduce computational load, we retained only five simulations for further analysis.}. We  selected the maximum scale to be around 500 kpc and 1 Mpc, in our case $\Lambda_{\rm max} = 470$ and 940 kpc, which means dividing the cube by $\rm k_{\rm min}=6$ and 3, respectively. From \citealt{DF_2019}, we know that the MF power spectrum can be more easily described with a single power law at scales smaller than the peak. For this reason, we  also tested  50 kpc as $\Lambda_{\rm max}$ ($\rm k_{\rm min}=56$). Moreover, since we do not have knowledge of the location of the radio galaxies along the LOS, we  adopted two versions of the RM map: for the sources embedded in the cluster (e.g. T1 and T2), we used the one delivered by the simulations (i.e. the one obtained from the integration along the $z$-direction of half the MF and gas density cubes starting from the centre).  The other one is produced through the integration of the entire cubes along the LOS and is used for the background sources (e.g. W1, W2, and Group G). In this way, we accounted for the different amount of intervening material along the $z$-direction, considering a proper location of the sources on the LOS. The 2D RM maps were then convolved with a Gaussian function with the same FWHM as the restoring beam of the observations (i.e. 33.4 kpc) and re-binned following the imaging procedure (see Sect. \ref{IMprep}).

\subsection{Constraining the maximum MF scale ($\Lambda_{\rm max}$)} \label{MF scale}
To properly compare the observed and simulated quantities, we  trimmed the observed RM map at 414 pixels according to the dimension of the simulated ones and created a mask from it. The latter was used to blank the ten simulated RM maps for each MF model and integration following the shape of the sources and the corresponding $\sigma_{\mathrm{RM}}$ was then extracted.
We  repeated this method for all the combinations of parameters considered in the previous section and calculated the reduced $\chi^2$ for $\sigma_{\mathrm{RM}}$ using the formula from \citealt{Govoni_2006}, as follows: 
\begin{equation}
    \chi^2_{\sigma_{\mathrm{RM}},red}=\dfrac{1}{d.o.f.}\displaystyle\sum^5_{i=1}\dfrac{(\sigma_{\mathrm{RM}obs,i}-|\sigma_{\mathrm{RM}}|_{sim,i})^2}{|\mathrm{Scatter}_{sim,i}|^2+\mathrm{Err}_{\sigma_{\mathrm{RM}}obs,i}^2},
    \label{eq:chi-squared}
\end{equation}
where $d.o.f.=3$ are the degrees of freedom of the model, namely, the number of data points (5) subtracted by the number of free parameters (2); $\sigma_{\mathrm{RM}obs,i}$ is the observed RM dispersion of every source (here: 5); and $\mathrm{Err}_{\sigma_{\mathrm{RM}}obs,i}$ is the corresponding error calculated as Eq. \ref{eq:RMerr}. Then, $|\sigma_{\mathrm{RM}}|_{sim,i}$ is the mean over five simulations of the RM dispersion extracted from that specific source. In the end, $|\mathrm{Scatter}_{sim,i}|$ is the mean of the scatter of five simulations in the annulus in which the source is located. Again, $|\sigma_{\mathrm{RM}}|_{sim,i}$ and $|\mathrm{Scatter}_{sim,i}|$ are extracted from the half-integration and the total-integration RM maps for the embedded and the background radio galaxies, respectively. The aim was to find what combination of parameters minimises the reduced $\chi^2$ value. 

\begin{figure}[!htb]
    \vspace{-0.45cm}
    \centering
    \includegraphics[width=0.8\linewidth]{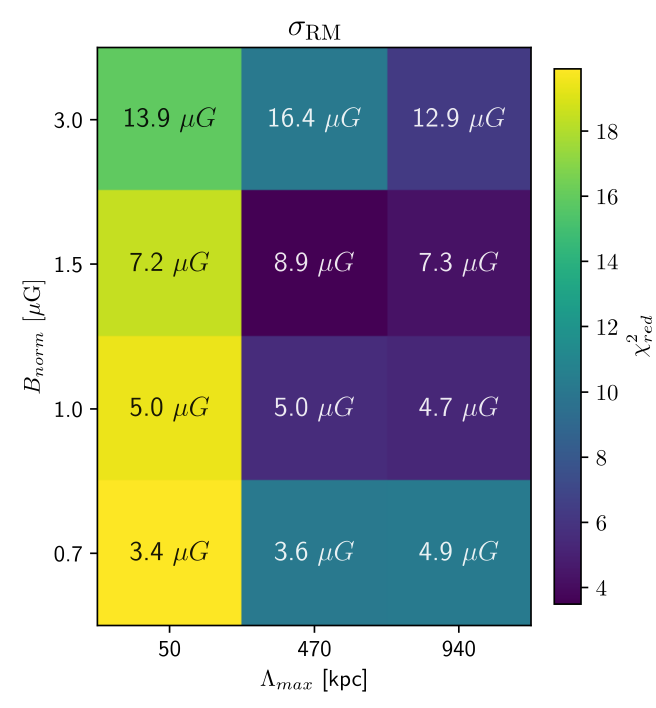}
    \caption{Reduced $\chi^2$ value of $\sigma_{\mathrm{RM}}$ for each combination of $\rm B_{\rm norm}$ and $\Lambda_{\rm max}$. The corresponding average central value of the MF, $\rm \langle B_0 \rangle$, is reported at the centre of each square. The minimum is 3.49 with $\rm \langle B_0 \rangle=8.9 \ \mu$G.}
    \label{fig:chi2-sigmaRM}
\end{figure}

We report the results in Fig. \ref{fig:chi2-sigmaRM} showing that the minimum of the reduced $\chi^2$ (3.49) was reached for $\Lambda_{\rm max}=470 \ \mathrm{kpc}$ and $\rm B_{\rm norm}=1.5 \ \mu$G, with mean central MF, $\rm \langle B_0 \rangle$, of $8.9 \ \pm \ 3.2 \ \mu$G. To compute $\langle \mathrm{B_0} \rangle$, we  created a new cube, with the same dimension of the other MF components cubes, but this time with the MF module, $|\vec{\mathbf{B}}|=\sqrt{3}|\vec{\mathbf{B_x}}|$, in each pixel. Then, the average MF module is extracted from spherical shells with increasing radius from the centre in order to create the MF profile (see Sect. \ref{MF peak}). $\rm \langle B_0 \rangle$ of every combination of parameters refers to the mean over five simulations in the first spherical shell (i.e. up to $\sim 70$ kpc) and $\mathrm{err}_{\mathrm{B_0}}$ is computed as the standard deviation.

\begin{figure}[!htb]
    \centering
    \includegraphics[width=0.95\linewidth]{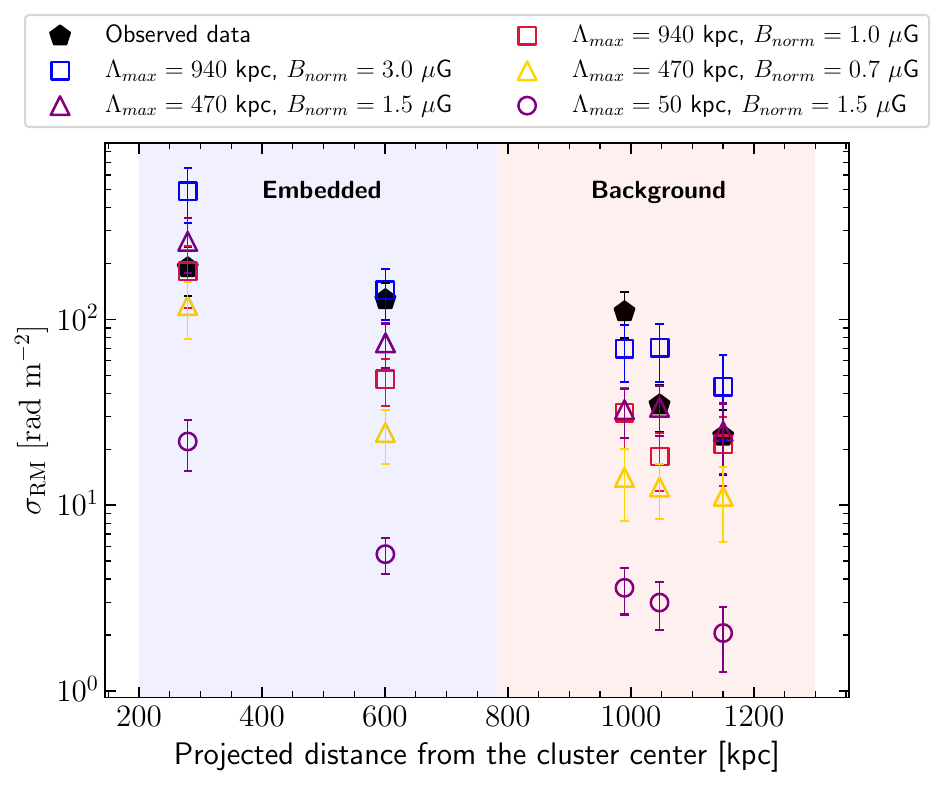}
    \caption{Comparison between the observed and the simulated $\sigma_{\mathrm{RM}}$ as a function of the projected distance from the cluster centre for the five radio galaxies analysed in this section (in distance order: T1, T2, W2, W1, Group G). The black pentagons refer to the observed data, where the errors are computed as in Eq. \ref{eq:RMerr}. Coloured symbols represent the mean over five simulations of $\sigma_{\mathrm{RM}}$ extracted from each source. From higher to lower: blue ($3.0 \ \mu$G), purple ($1.5 \ \mu$G), bordeaux ($1.0 \ \mu$G), and yellow ($0.7 \ \mu$G). Different shapes represent different maximum fluctuating spatial scales, $\Lambda_{\rm max}$, of the MF. From higher to lower: square (940 kpc), triangle (470 kpc), and circle (50 kpc). Again, the simulated values are taken from the half-integration RM maps for the embedded sources, whereas for the background
radio galaxies, they were taken from the total-integration
maps. The associated error is computed as the sum in quadrature of Eq. \ref{eq:RMerr} and the mean of the scatter of five simulations in the annulus in which the source is located. The $\sigma_{\mathrm{RM}}$ axis is in logarithmic scale for a better visualisation of the profile.}
    \label{fig:mean_sim_total_pp}
\end{figure}
We  compared the observed and the simulated $\sigma_{\mathrm{RM}}$ as a function of the projected distance from the cluster centre for the five radio galaxies analysed in this section. Examples are reported together in Fig. \ref{fig:mean_sim_total_pp} for different MF models. The MF model responsible for the minimum in the reduced $\chi^2$ is the  one that best matches the observations. However, the most evident feature is the departure of the third source in distance from the centre (W2) with respect to the simulated points. By performing the same analysis as in Sect. \ref{RM Synthesis Results} for the FDF spectrum of W2 (Fig. \ref{subfig:FDF-W2}), we found evidence of multiple components, but not a real double peak as in T1. There is a possible contribution from the superimposed Galactic bubble, increasing $\sigma_{\mathrm{RM}}$ for W2, that we are not able to disentangle with the current resolution in Faraday space. After removing W2 from the statistical analysis, we noticed some changes in the reduced $\chi^2$: the statistic worsens in the case of all that models with very high $\chi^2_{red}$ values ($> 6$); for instance, those with $\Lambda_{\rm max}=50$ kpc, since in those cases getting rid of one point becomes determinant. On the other hand, the statistical results improve to 2.09 and 3.48 for the models with $\Lambda_{\rm max}=470$ kpc or 940 kpc and $\rm B_{\rm norm}=1.5 \ \mu$G, respectively, as expected for the reasoning mentioned above. The relevant conclusion is that, even without considering this source, the minimum does not change. The best reproducing model remains the same, with a $\chi^2_{red}$ of 2.09.
It is clear that none of the simulated outcomes perfectly fits the observed values, suggesting that our model is too simplistic for the observations we are analysing. Ultimately, we found that we are missing sources located in the very central regions of the cluster ($\lesssim 250 \ \mathrm{kpc}$), which would otherwise have helped in finding the best-fit profile.

\subsection{Constraining the peak of the MF power spectrum ($\Lambda_{\rm peak}$)} \label{MF peak}
As the minimum $\chi^2_{red}$ found in the previous analysis is high and the MF power spectrum was chosen among a small set of simulated clusters, in this section, we explore other MF power spectrum peaks: $\Lambda_{\rm peak}$,  mainly at $\sim 70$ and $\sim 140$ kpc, setting the parameter C to $20.1$ and $10.1 \ \mathrm{Mpc^{-1}}$, respectively. In this case, we would consider MF models which are not present in the simulations by \citealt{DF_2019}. The main goal of this step is to understand whether or not a different MF model better reproduces the observed data and consequently delivers improved statistical results. 

We followed the same steps presented in Sects. \ref{MIROcode} and \ref{MF scale}, this time fixing $\Lambda_{\rm max}$ at 470 kpc, as obtained from the profile with the minimum $\chi^2_{red}$, and changing $\rm B_{\rm norm}$ and C in the set of values [1.0, 1.5, 2.0, 3.0] $\mu$G and [20.1, 10.1, 5.0] $\mathrm{Mpc^{-1}}$, respectively\footnote{Some of the models, e.g. with $\rm C = 5.0 \ \mathrm{Mpc^{-1}}$ and $\rm B_{\rm norm} = [1.0, 1.5, 3.0] \ \mu$G are simulated and analysed in the previous section. Here, the corresponding outcomes are used for completeness in the comparison with the new models.}. Therefore, the number of free parameters is again 2 (i.e. 3 d.o.f. in Eq. \ref{eq:chi-squared}). All the other quantities reported in Table \ref{tab:Sim specifics} remain the same. 

\begin{figure}[!htb]
    \centering
    \includegraphics[width=0.8\linewidth]{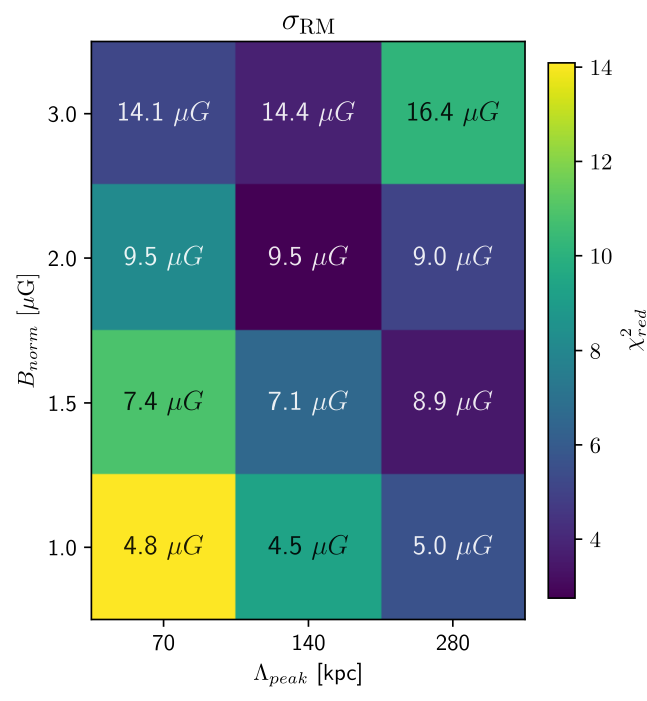}
    \caption{Reduced $\chi^2$ value of $\sigma_{\mathrm{RM}}$ for each combination of $\rm B_{\rm norm}$ and $\Lambda_{\rm peak}$. The corresponding average central value of the MF, $\rm \langle B_0 \rangle$, is reported at the centre of each square. The minimum is 2.75 with $\rm \langle B_0 \rangle=9.5 \ \mu$G.}
    \label{fig:chi2-sigmaRM-new}
\end{figure}

\begin{figure}[!htb]
    \centering
    \includegraphics[width=0.95\linewidth]{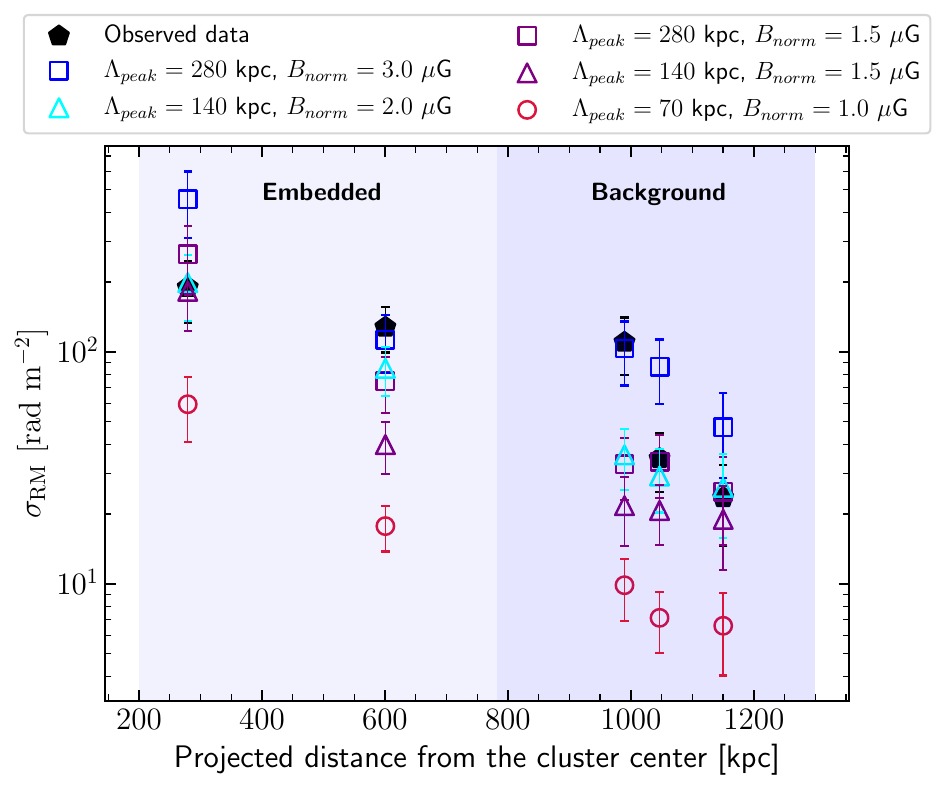}
    \caption{Comparison between the observed and the simulated $\sigma_{\mathrm{RM}}$ as a function of the projected distance from the cluster centre for the five radio galaxies analysed in this Section (in distance order: T1, T2, W2, W1, Group G). The black pentagons refer to the observed data, where the errors are computed as in Eq. \ref{eq:RMerr}. Coloured symbols represent the mean over five simulations of $\sigma_{\mathrm{RM}}$ extracted from each source. From higher to lower:  blue ($3.0 \ \mu$G), light-blue ($2.0 \ \mu$G), purple ($1.5 \ \mu$G), and bordeaux ($1.0 \ \mu$G). Different shapes represent different MF power spectrum peaks. From higher to lower: square (280 kpc), triangle (140 kpc), and circle (70 kpc). Again, the simulated values are taken from the half-integration RM maps for the embedded sources, whereas for the background radio galaxies, they were taken from the total-integration
maps. The associated error is computed as the sum in quadrature of Eq. \ref{eq:RMerr} and the mean of the scatter of five simulations in the annulus in which the source is located. The $\sigma_{\mathrm{RM}}$ axis is in logarithmic scale for a better visualisation of the profile.}
    \label{fig:mean_sim_new_pp}
\end{figure}

\begin{figure*}
    \begin{subfigure}{0.485\linewidth}
        \centering
        \includegraphics[width=0.92\linewidth]{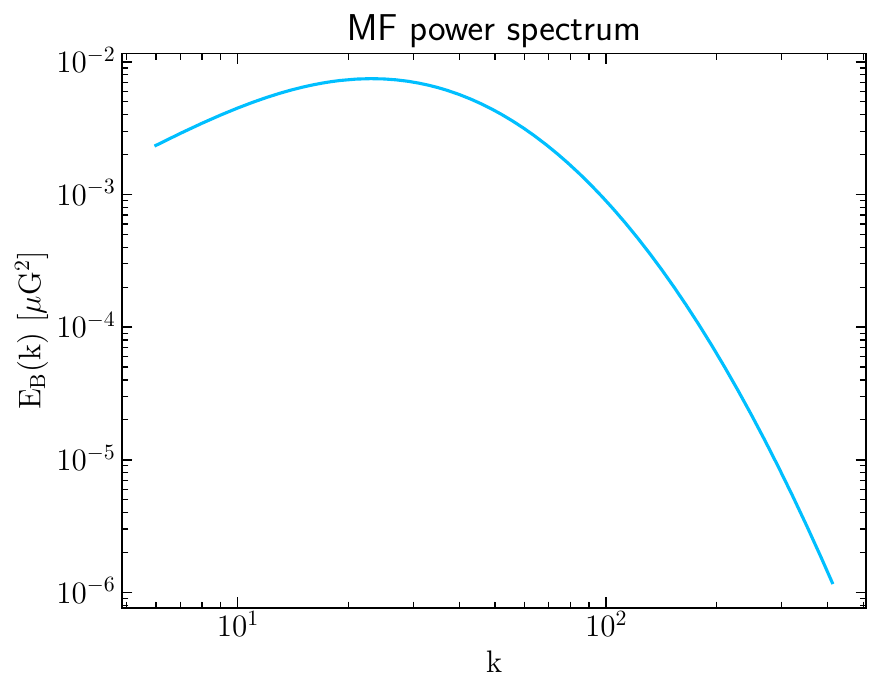} 
        \caption{}
        \label{subfig:MFS-best-fit}
    \end{subfigure}
    \begin{subfigure}{0.53\linewidth}
        \centering
        \includegraphics[width=0.92\linewidth]{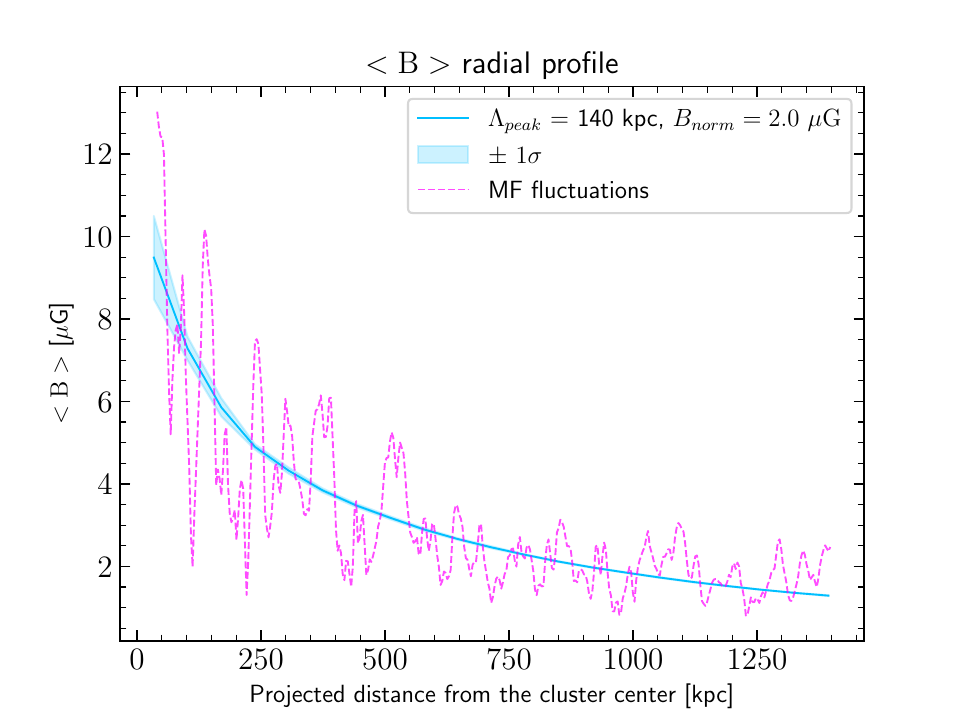} 
        \caption{}
        \label{subfig:B_best-fit}
    \end{subfigure}
    \caption{Panel (a): Best-fit MF power spectrum, ranging between $\rm \Lambda_{min}=7 \ \mathrm{kpc}$ ($\rm k_{max}=413$) and $\Lambda_{\rm max} = 470 \ \mathrm{kpc}$ ($\rm k_{min}=6$), with peak at $\sim$140 kpc ($\rm C=10.1 \ \mathrm{Mpc^{-1}}$, $\rm k \sim 20$). Panel (b): Profile of $\langle\mathrm{B}\rangle$ as a function of the projected distance from the cluster centre. It is the mean over five simulations and the scatter is the standard deviation. The central average MF, $\langle\mathrm{B_0}\rangle$, is equal to $9.5 \pm 1.0 \ \mu$G. Dashed pink line represents the MF fluctuations taken along the central LOS.}
    \label{fig:best-fit}
\end{figure*}

We report the results of the reduced $\chi^2$ for $\sigma_{\mathrm{RM}}$ in Fig. \ref{fig:chi2-sigmaRM-new}. With the new models, the minimum of the reduced $\chi^2$ (2.75) is reached for $\Lambda_{\rm max}=470 \ \mathrm{kpc}$ and $\rm B_{\rm norm}=2.0 \ \mu$G, but with $\Lambda_{\rm peak}$ of $\sim 140$ kpc and mean central MF, $\rm \langle B_0 \rangle$, of $9.5 \ \pm \ 1.0 \ \mu$G. The latter result is consistent with the mean central MF obtained for the best-fit model found in the initial analysis. The improvement in the statistical outcome could be attributed to a larger scatter in the simulations, as well as a better agreement between the observed and simulated data of the radio galaxies. We show this comparison again in Fig. \ref{fig:mean_sim_new_pp}. Indeed, the light blue triangles ($\Lambda_{\rm peak}$ $= 140$ kpc, $\rm B_{\rm norm}=2.0 \ \mu$G) better reproduce  the observed data (black pentagons) within the errors with respect to the purple squares ($\Lambda_{\rm peak}$ $= 280$ kpc, $\rm B_{\rm norm}=1.5 \ \mu$G), especially for the two embedded sources (T1 and T2). Finally, also in this case, the minimum does not change without considering W2 in the statistical analysis, and $\chi^2_{red}$ of the best-fit model improves from 2.75 to 1.24. \\
\\
We show in Fig. \ref{subfig:MFS-best-fit} the best-fit MF power spectrum inserted in the simulations, following Eq. \ref{eq:PowerSpectrum} and ranging between $\rm \Lambda_{min}=7 \ \mathrm{kpc}$ and $\Lambda_{\rm max} = 470 \ \mathrm{kpc}$, with a peak at 140 kpc. The corresponding best-fit MF intensity profile (Fig. \ref{subfig:B_best-fit}) was computed as the mean over the five simulations and the scatter is again the standard deviation. Each profile was obtained from the MF module cube, as explained in Sect. \ref{MF scale}.

\section{Discussion} \label{Discussion}
Starting from the results of the RM analysis, the aim of this work is to constrain A2142 MF intensity, profile,  and power spectrum from the simulated MF model that may reproduce better the observed $\sigma_{\mathrm{RM}}$ radial profile. We found that $\sigma_{\mathrm{RM}}$ decreases with increasing projected distance from the cluster centre, with values going from several hundreds in the core to tens of $\mathrm{rad \ m^{-2}}$ in the peripheral regions, suggesting that the MF is particularly strong and turbulent centrally and becomes less intense towards larger distances. These are in line with other results found in literature (see e.g. \citealt{Bonafede_2010}, \citeyear{Bonafede_2013}; \citealt{Vacca_2010}; \citealt{Stuardi_2021}; \citealt{Osinga_2024}). The lower values in $\sigma_{\mathrm{RM}}$ of W1 and Group G with respect to the general trend could be possibly due to fluctuations in the MF spectrum (see Fig. \ref{subfig:B_best-fit}). They could be located in a region where the MF intensity is lower with respect to the average one at that projected distance, producing smaller dispersion in the RM. This demonstrates the power of detecting multiple sources at the same projected distance. 

Through the comparison with simulations, we found that a MF tangled on scales between 7 and 470 kpc, following a power spectrum with a peak at $\sim 140$ kpc, best describes our data, with a mean central MF of $9.5 \ \pm \ 1.0 \ \mu$G and $\eta=0.5$. The comparison with previous works on other clusters is presented in Sect. \ref{Comparison}, whereas in Sect. \ref{Limitations} we discuss the limitations of this model and the possible future perspectives. In Sect. \ref{originH1} we explore the implications of our results for understanding the origin of the diffuse radio emission observed in A2142.

\subsection{Comparison with MF estimates in other clusters} \label{Comparison}
Information about the MF in individual galaxy clusters has been obtained thus far via RM studies only for few objects, while adopting different approaches. 

In our work, we assumed a thermal electron density profile described by Eq. \ref{eq:densityprofile}, which was obtained from the fitting of the de-projected density profiles obtained from X-ray observations of NCC clusters, including our target. Moreover, we assumed a MF power spectrum described by Eq. \ref{eq:PowerSpectrum}, obtained from MHD cosmological simulations. This was also adopted in the past to constrain the MF profile and intensity in merging clusters (\citealt{Stuardi_2021}; \citealt{DeRubeis_2024}), whereas other works assumed a Kolmogorov spectrum with different power law indexes (e.g. \citealt{Murgia_2004}; \citealt{Guidetti_2008}; \citealt{Govoni_2006}, \citeyear{Govoni_2017}; \citealt{Bonafede_2010}; \citealt{Vacca_2010}, \citeyear{Vacca_2012}).

A2142 is the first object for which the MF was constrained using this approach in a system that is neither in a relaxed dynamical state nor undergoing a major merger, being classified as an intermediate case between a CC and a NCC cluster. In our study, we found an average central MF, $\rm \langle B_0 \rangle$, of $9.5 \ \mu$G, which is consistent with the values found in previous works, which range between 1.3 and 11.7 $\mu$G. A2142 is comparable in mass with the merging cluster Abell 665 \citep{Vacca_2010}, which was found to have a lower central MF of $1.3 \ \mu$G. There is only one cluster with central MF higher than A2142 and it is a CC cluster \citep[A2199,][]{Vacca_2012}. However, as pointed out by \citealt{Stuardi_2021}, there is no evident correlation between $\rm \langle B_0 \rangle$ and the mass of the objects, and the statistics are not yet sufficient to determine whether or not relaxed clusters present higher central MF intensities with respect to merging ones. 

We  also computed the average MF strength in 1 $\mathrm{Mpc^{3}}$ volume, $\rm \langle B_{1\mathrm{Mpc^3}}\rangle \sim 3.5 \ \mu$G, from the MF module cube, within a sphere with radius of 620 kpc, namely, $\rm V=4\pi r^3/3  \sim 1 \ \mathrm{Mpc^3}$, and by taking the mean over five simulations. It is slightly higher with respect to the values found in literature (e.g. between 0.2 and 2 $\mu$G), but A2142 shows evidence of possible minor mergers during its evolution (\citealt{Owers_2011}), which may contribute to the MF amplification. Moreover, the average MF within 1.2 Mpc (i.e. the size of H3) is $\sim 2.1 \ \mu$G and is larger than the MF equipartition estimate of $\sim 0.2 \ \mu$G, which was derived with Eq. 26 of \citealt{Govoni_2004} and using the parameters for the radio halo H3 from Table 2 of \citealt{Riseley_2024}. 
This is not surprising, as RM-derived values are systematically higher than those obtained from equipartition estimates based on radio halo emission (e.g. \citealt{VanWeeren_2009}; \citealt{Vacca_2010}; \citealt{Bonafede_2010}). The latter represent volume-averaged values and depend on several underlying assumptions. 

\subsection{Magnetic field and origin of the diffuse radio emission}\label{originH1}
A2142 hosts a central halo component (H1), which was originally classified as a mini-halo \citep{Venturi_2017}. The hadronic or leptonic origin of clusters' radio sources is a widely-debated field and many authors have investigated the connection between thermal and non-thermal components to understand the origin of radio mini-halos (e.g. \citealt{Bravi_2016}; \citealt{Giantucci_2019}).

Recently, \citealt{Ignesti_2020}  performed this study in a sample of radio mini-halos found in relaxed clusters using radio/X-ray surface brightness correlations to constrain the physical parameters of a hadronic model.  They have assumed stationary conditions, without including the effect of the re-acceleration. They  found a super-linear correlation between the radio and X-ray surface brightness. This suggests a peaked distribution of relativistic electrons and a MF intensity in the range of $10-40 \ \mu \rm G$ for $\eta=0.5$. In contrast, H1 shows a sub-linear correlation (\citealt{Bruno_2023}; \citealt{Riseley_2024}), suggesting a broader distribution of relativistic electrons, potentially re-accelerated by merger events in the evolutionary history of A2142. The average central MF strength derived in our study is at the low end of the values predicted by the hadronic model of  \citealt{Ignesti_2020}. However, the average MF intensity of $6.4 \ \mu$G within 200 kpc (computed in the same way as above) satisfies the lower limit of $4 \ \mu$G, which is consistent with a possible hadronic origin of H1 in A2142 \citep{Bruno_2023}.
In conclusion, a hadronic origin for H1 cannot be ruled out and it is plausible that both hadronic and re-acceleration processes contribute to its emission.

\subsection{Limitation of the MF modelling} \label{Limitations}
To reduce the number of free parameters in the MF modelling, the MF strength is assumed to decrease as the square root of the thermal electron density (Eq. \ref{eq:etaB}) as found by \citealt{Bonafede_2010} and several previous authors. Subsequent works have obtained higher values of $\eta$ (e.g. between 0.9 and 1.1), which imply both a higher $\rm \langle B_0 \rangle$ and a steeper profile of the MF energy density. In this case, fixing this index breaks the degeneracy between these two quantities. We allowed the maximum fluctuating spatial scale, $\rm \Lambda_{\rm max}$ (i.e. $\rm k_{min}$) to change, but not the minimum one, $\rm \Lambda_{min}$ (i.e. $\rm k_{max}$).  
In our case, the code set $\rm k_{max}$ as the maximum possible Fourier scale in the cube (i.e. $\sim 413$) and this translates in the minimum possible spatial scale of $\sim$ 7 kpc. This value is almost five times smaller than our resolution; therefore, we fixed it to reduce the number of free parameters. The introduction of smaller fluctuations would require a decrease in the pixel scale. This would produce further inhomogeneities in the RM distribution that  cannot be resolved, since the RM maps have been smoothed to 33.4 kpc in resolution.

Regarding the nature of the cluster, we assumed a post-merger dynamical state for the MF power spectrum (ID E1 of \citealt{DF_2019}) and explored other shapes with a different peak by changing the parameter C. We found a more statistically significant model (reduced $\chi^{2} = 2.75$)\footnote{which is improved to 1.24, once removing W2 from the comparison.} with $\rm B_{\rm norm}= 2 \ \mu$G and $\Lambda_{\rm peak}$ $\sim 140$ kpc. The latter value is consistent with the maximum fluctuating spatial scales found by authors adopting a Kolmogorov power spectrum as MF model (e.g. \citealt{Murgia_2004}; \citealt{Guidetti_2008}; \citealt{Govoni_2006}, \citeyear{Govoni_2017}; \citealt{Bonafede_2010}; \citealt{Vacca_2010}, \citeyear{Vacca_2012}).  Overall, these results are consistent with those obtained from other clusters (\citealt{Murgia_2004}; \citealt{Guidetti_2008}; \citealt{Govoni_2006}, \citeyear{Govoni_2017}; \citealt{Bonafede_2010}; \citealt{Vacca_2010}, \citeyear{Vacca_2012}; \citealt{Stuardi_2021}; \citealt{DeRubeis_2024}), even though none of the simulated outcomes perfectly fit the observed values, suggesting that our model is too simplistic for the cluster we are analysing.

Cluster radio galaxies interact with the surrounding medium and are capable of reshaping locally the MF and density distribution. This may cause a bias in the RM and amount of depolarisation. \citealt{Osinga_2022} did not find any significant difference between the depolarisation of cluster members and background sources in a statistical study of 124 galaxy clusters. The observed radial decline of $\sigma_{\mathrm{RM}}$ supports the fact that the Faraday rotation is caused by the ICM, whereas the RM arising locally to the embedded radio sources can be neglected. Future observations with a greater angular resolution would allow for the  beam depolarisation to be minimised and for a higher polarised signal to be detected, especially from the sources in the very central regions of the cluster. A broader observational bandwidth would improve the FWHM of the RMTF, thereby  helping to resolve the narrow multiple peaks from components on the same LOS. Finally, we note that the spherical symmetry assumption may not reproduce the real distribution of the ICM in A2142, as it is elongated in a particular direction due to a past merger event. 

\section{Conclusions} \label{Conclusion}
The aim of this work is to study and constrain the MF intensity, profile, and power spectrum in A2142. This galaxy cluster is peculiar from its dynamical point of view, as it is considered a warm-cool-core cluster, representing an additional case for understanding the MF amplification level in such systems.

We present  high-sensitivity MeerKAT L-band ($872-1712$ MHz) data, which have been imaged in polarisation for the first time in order to study the MF intensity and profile in A2142. We created cubes of Stokes Q and U maps in frequency, convolved to a common resolution of $20^{\prime\prime}\times20^{\prime\prime}$, and applied the RM synthesis technique. We  produced the RM and the fractional polarisation, $\mathrm{{F_p}}$, maps from the RM synthesis outputs and analysed the average RM, $\langle \mathrm{RM} \rangle$, RM dispersion, $\sigma_{\mathrm{RM}}$, and $\mathrm{{F_p}}$ as a function of the projected distance from the cluster centre. From the $\langle \mathrm{RM} \rangle$ and the $\sigma_{\mathrm{RM}}$ radial profiles, we conclude that the MF becomes less intense toward larger distances. The ${\mathrm{F_p}}$ radial profile confirms that radio galaxies seen in projection near the cluster centre suffer higher depolarisation than peripheral ones. The very low values of $\mathrm{{F_p}}$ ($<1\%$) found for the central radio galaxy, T1, are a direct consequence of multiple emitting and rotating materials which spread the polarised emission at different Faraday depth.

Starting from these results, we built 3D simulations using a modified version of the \texttt{MIRO'} code (\citealt{Bonafede_2013}; \citealt{Stuardi_2021}) in order to produce mock RM maps. We assumed a 3D gas density model, following the electron density profile for NCC clusters (including A2142) found by \citealt{Ghirardini_2019} and a 3D MF model from the power spectrum for post-merger clusters (ID E1) derived by recent cosmological MHD simulations (\citealt{DF_2019}). As a first analysis, we  fixed four of the six parameters available ($\eta$, $\rm \Lambda_{min}$, B and C) and left the other two ($\Lambda_{\rm max}$ and $\rm B_{\rm norm}$) to vary in a defined range of values. In a second moment, we explored different shapes of the MF power spectrum by changing the peak, $\Lambda_{\rm peak}$. We  compared the simulated outcomes with the observational results.

We computed the reduced $\chi^2$ comparing simulated and observed $\sigma_{\mathrm{RM}}$ to find the best-fit parameters of the MF model. We found that a MF tangled on scales between 7 and 470 kpc, following a power spectrum with a peak at $\sim 140$ kpc, best describes our data with a mean central MF of $9.5 \ \pm \ 1.0 \ \mu$G and $\eta=0.5$. Only one radio galaxy, W2, presents a clear departure once compared to the simulated $\sigma_{\mathrm{RM}}$ extracted from that location. The average central MF intensity computed from this model does not rule out a hadronic origin of the radio halo component H1, in A2142. Based on a comparison of our results with the literature, we find a good agreement, despite the different approaches used and the dynamical state of the galaxy clusters. 

A2142 is the first object observed with the SKA precursor MeerKAT interferometer for which the MF has been constrained with state-of-the-art techniques, such as the RM synthesis technique, while  adopting the most recent results from X-ray observations and simulations. Even though this kind of study is not trivial, future works pursued in this field (both locally and at higher redshift) would give a more precise and realistic view of the MF and help improve our understanding of the non-thermal components in galaxy clusters.

\begin{acknowledgements}
The MeerKAT telescope is operated by the South African Radio Astronomy Observatory, which is a facility of the National Research Foundation, an agency of the Department of Science and Innovation.
This research made use of the LOFAR-IT computing infrastructure supported and operated by INAF, including the resources within the PLEIADI special `LOFAR' project by USC-C of INAF, and by the Physics Dept. of Turin University (under the agreement with Consorzio Interuniversitario per la Fisica Spaziale) at the C3S Supercomputing Centre, Italy. The data published here have been reduced using the CARACal pipeline, partially supported by ERC Starting grant number 679627 `FORNAX', MAECI Grant Number ZA18GR02, DST-NRF Grant Number 113121 as part of the ISARP Joint Research Scheme, and BMBF project 05A17PC2 for D-MeerKAT. Information about CARACal can be obtained online under the URL: \url{https://caracal.readthedocs.io}.
CJR acknowledges financial support from the German Science Foundation DFG, via the Collaborative Research Center SFB1491 `Cosmic Interacting Matters – From Source to Signal', and from the ERC Starting Grant `DRANOEL', number 714245. 
CS acknowledges support by the Fondazione ICSC, Spoke 3 Astrophysics and Cosmos Observations, National Recovery and Resilience Plan (Piano Nazionale di Ripresa e Resilienza, PNRR) Project ID $\rm CN\_00000013$ `Italian Research Center for High-Performance Computing, Big Data and Quantum Computing' funded by MUR Missione 4 Componente 2 Investimento 1.4: Potenziamento strutture di ricerca e creazione di `campioni nazionali di R$\&$ S (M4C2-19)' – Next Generation EU (NGEU).

This research has made use of several python packages not specifically cited in the text: \textsc{astropy}, a community-developed core Python package for Astronomy \citep{Astropy_2013,Astropy_2018,Astropy_2022}, \textsc{aplpy} \citep{Robitaille2012}, \textsc{matplotlib} \citep{Hunter2007}, \textsc{numpy} \citep{Numpy2011,Harris_2020_NumPy} and \textsc{scipy} \citep{Jones2001}. This research made extensive use of the Astrophysics Data System (ADS), funded by NASA under Cooperative Agreement 80NSSC21M00561.

\end{acknowledgements}

\bibliography{aa54881-25corr}
\bibliographystyle{aa}

\begin{appendix}
\onecolumn
\section{FDF spectrum, RM map, and $\mathrm{F_p}$ map of other radio galaxies in the field} \label{AppA}
\begin{figure}[!htb]
    \vspace{-0.1cm}
    \begin{subfigure}{0.475\linewidth}
        \centering
        \includegraphics[width=0.95\linewidth]{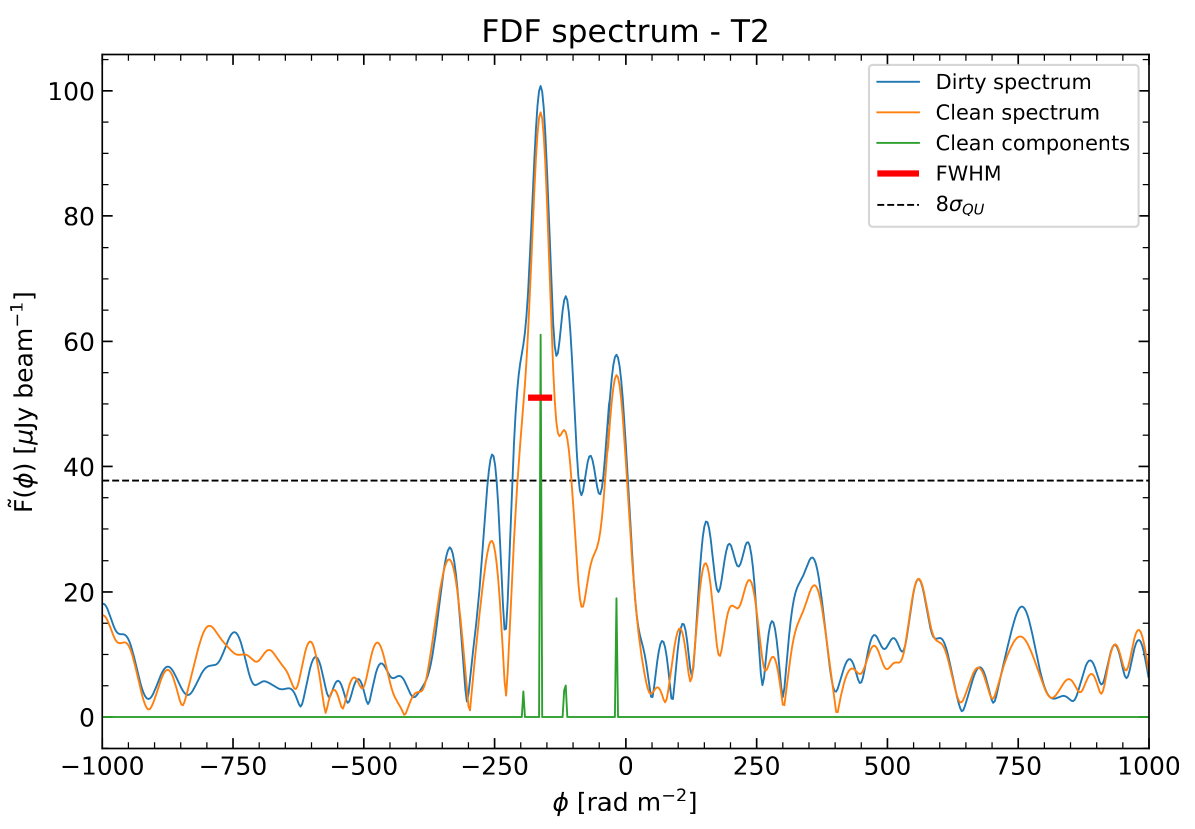}
        \caption{}
        \label{subfig:FDF-T2}
    \end{subfigure}
    \hspace{-0.5cm}
    \begin{subfigure}{0.475\linewidth}
        \centering
        \includegraphics[width=0.95\linewidth]{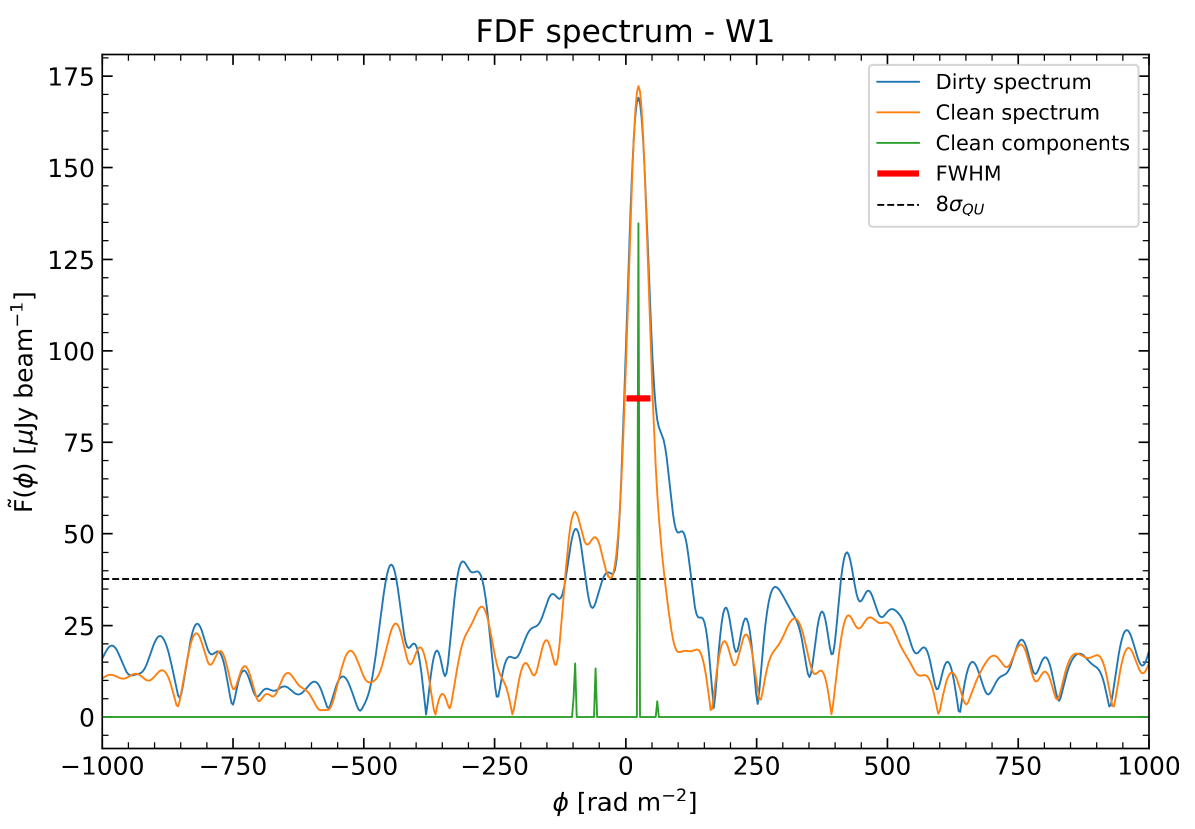}
        \caption{}
        \label{subfig:FDF-W1}
    \end{subfigure}
    \begin{subfigure}{0.475\linewidth}
        \centering
        \includegraphics[width=0.95\linewidth]{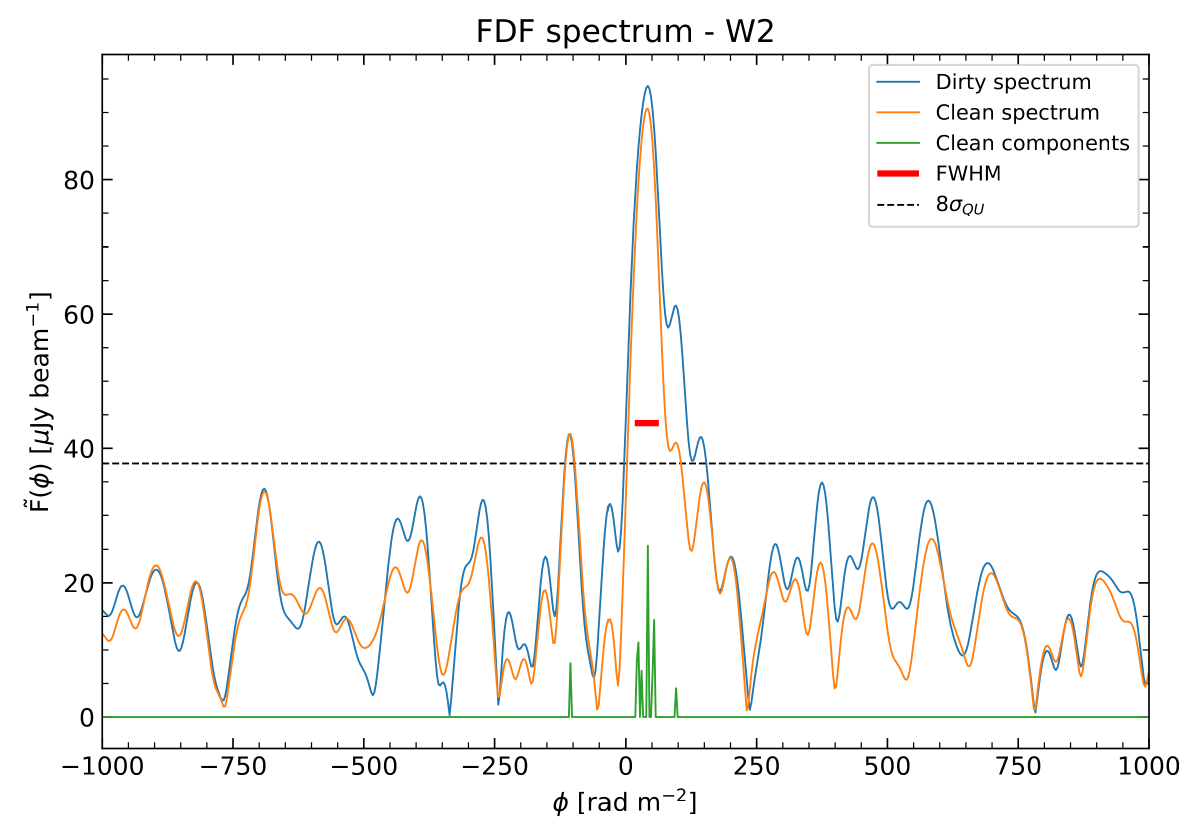}
        \caption{}
        \label{subfig:FDF-W2}
    \end{subfigure}
    \hspace{-0.5cm}
    \begin{subfigure}{0.475\linewidth}
        \centering
        \includegraphics[width=0.95\linewidth]{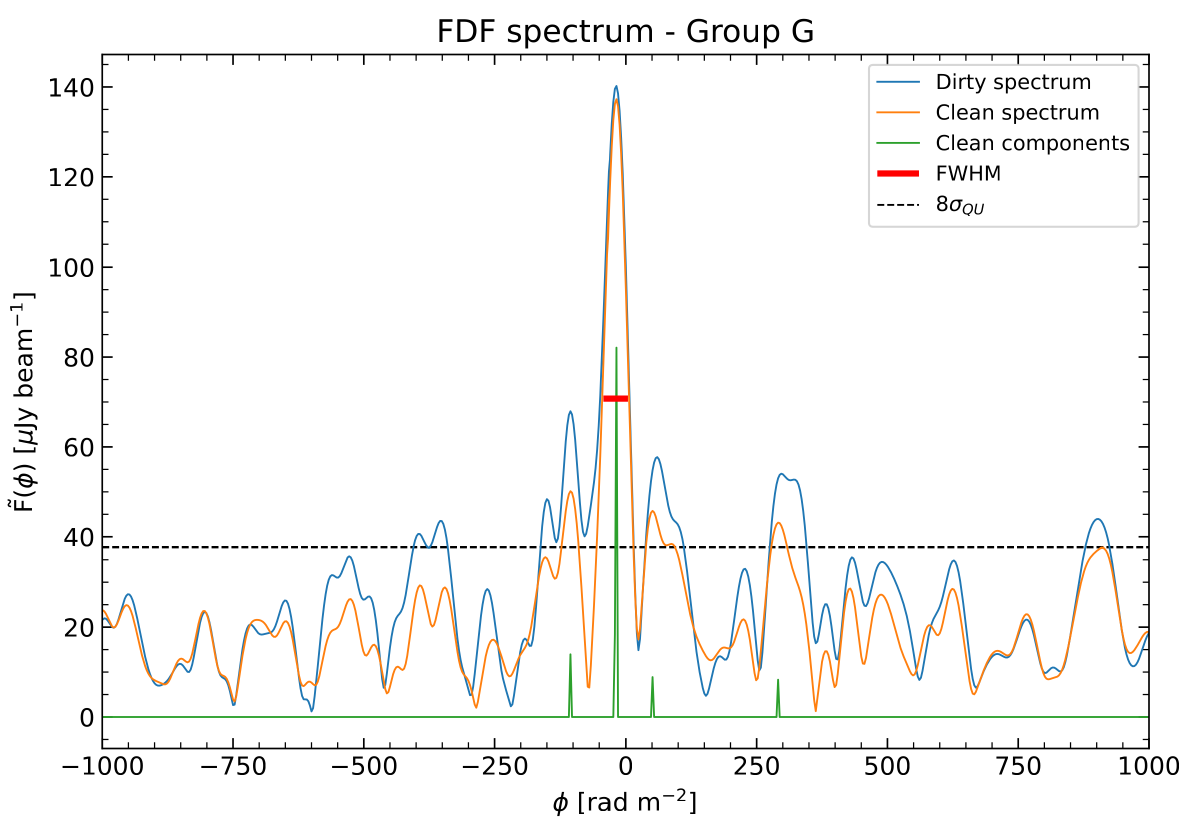} 
        \caption{}
        \label{subfig:FDF-GG}
    \end{subfigure}
    \begin{subfigure}{0.48\linewidth}
        \hspace{-0.4cm}
        \centering
        \includegraphics[width=0.965\linewidth]{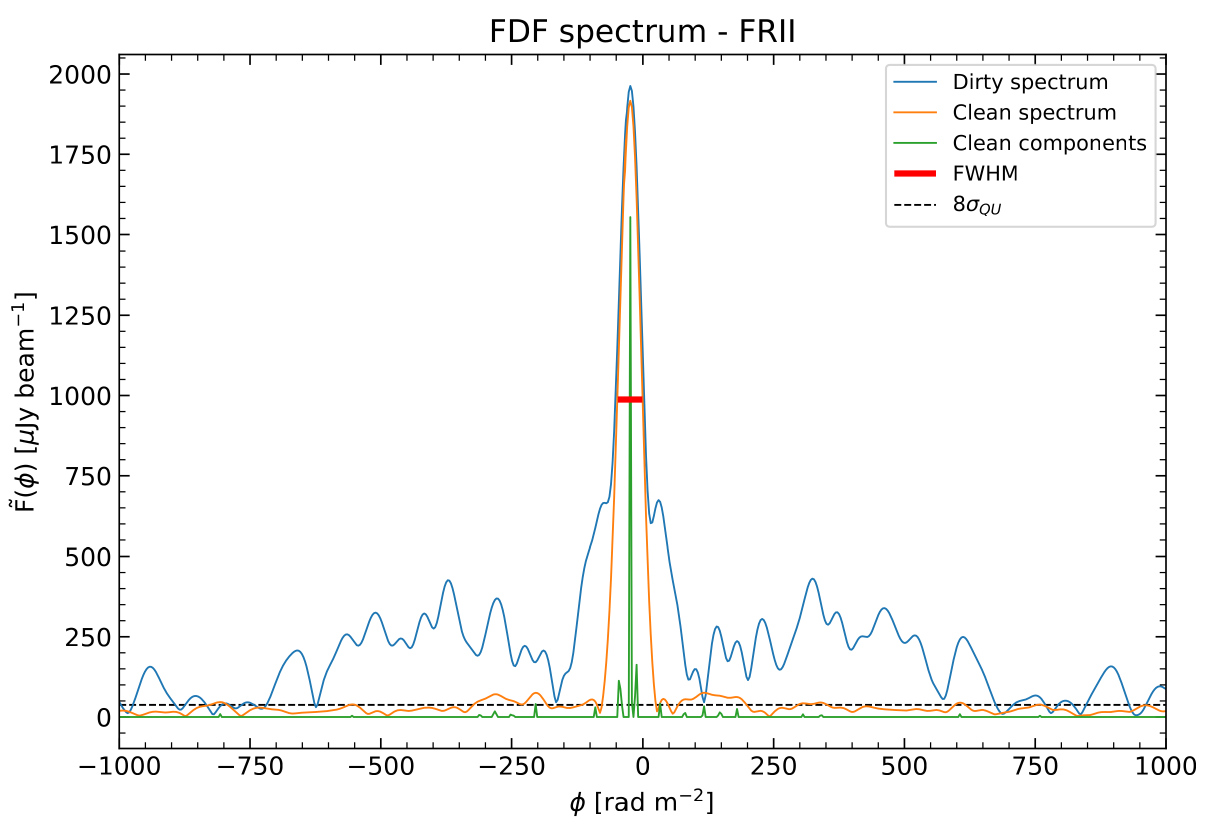} 
        \caption{}
        \label{subfig:FDF-FRII}
    \end{subfigure}
    \hspace{0.5cm}
    \begin{subfigure}{0.48\linewidth}
        \centering
        \includegraphics[width=0.965\linewidth]{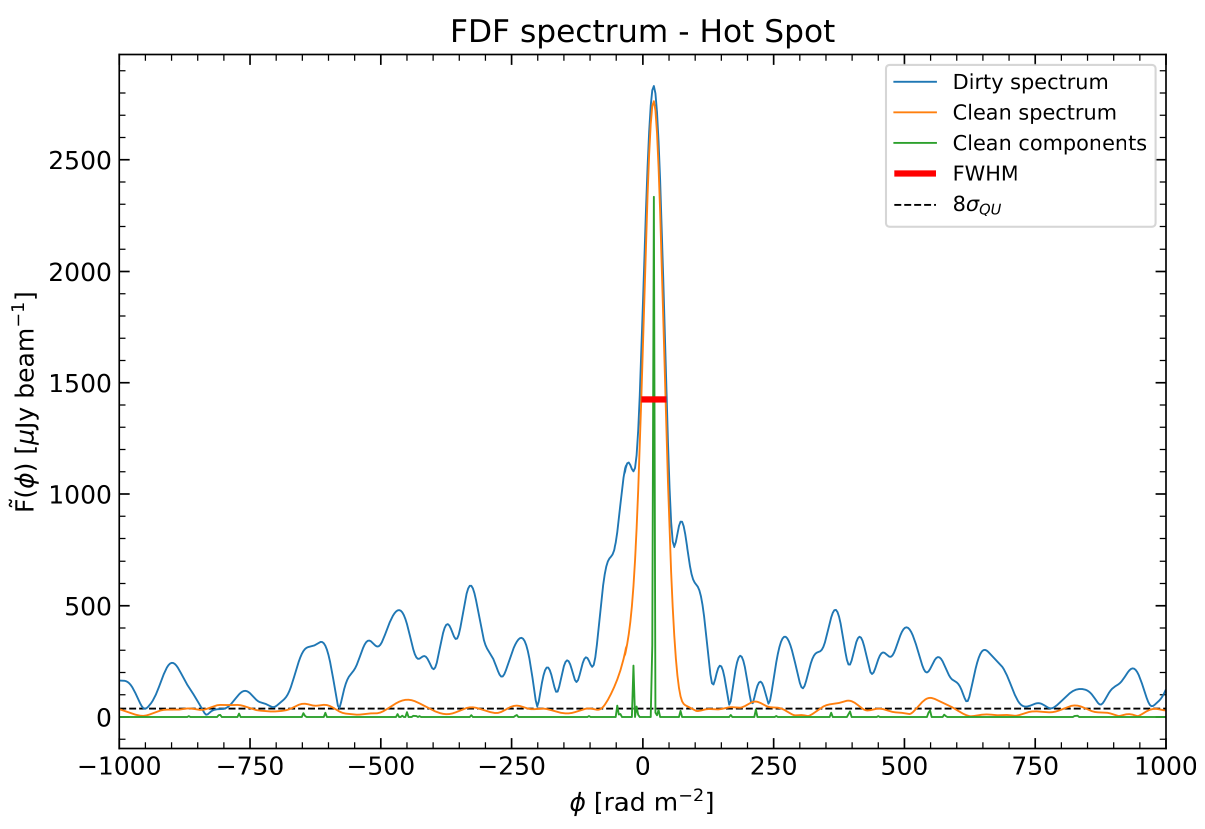} 
        \caption{}
        \label{subfig:FDF-HS}
    \end{subfigure}
    \caption{Reconstructed FDF spectrum taken from the brightest polarised pixel from some sources in the field. Panel (a): T2. Panel (b): W1. Panel (c): W2. Panel (d): Group G. Panel (e): FRII. Panel (f): Hot spot of a radio galaxy in the top of the image. The dirty spectrum is shown in blue, the clean spectrum in orange and the clean components in green. The FWHM of the RMTF is reported in red and the $8\sigma_\mathrm{{QU}}$ with the black dashed line as a reference.}
    \label{fig:other_FDF}
    \vspace{-2cm}
\end{figure}

\begin{figure}[!htb]
    \caption{Zoom of the RM map in correspondence of some sources analysed in this work with overlaid radio contours (1283 MHz), from $3\sigma_{\mathrm{I}}$ and scaling by a factor of 2. The $6\sigma_{\mathrm{QU}}$ and $3\sigma_{\mathrm{I}}$ detection thresholds were imposed in polarisation and in total intensity and only pixels above them are shown. Values were corrected for the Galactic foreground rotation. Panel (a): TL1, TL2, and TL3. Panel (b): TL4. Panel (c): BR1 and BR2. Panel (d): BL1. Panel (e): TR1. Positive values refer to the MF orientation along the LOS toward the observer, whereas negative values refer to an orientation away from the observer. Spatial scales and resolution beam are reported on the edges of the images.}
    \vspace{-0.2cm}
    \begin{subfigure}{0.51\linewidth}
        \centering
        \includegraphics[width=0.91\linewidth]{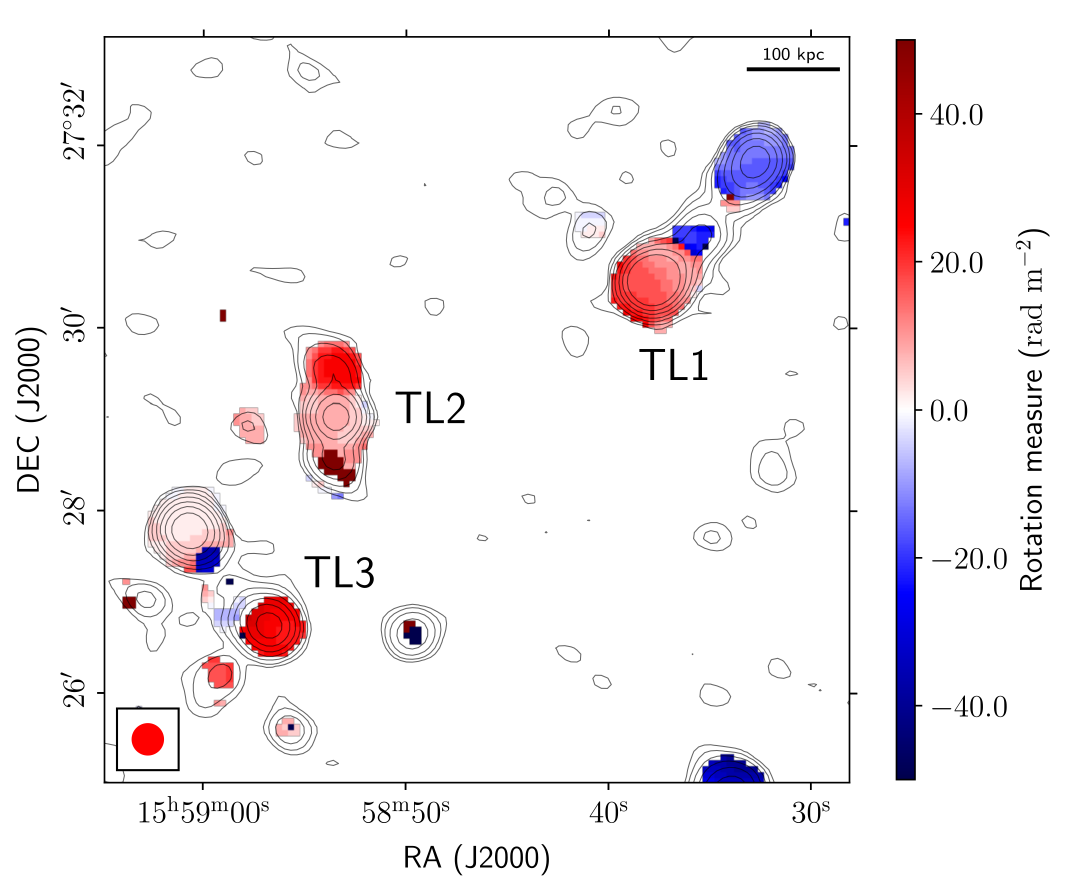}
        \caption{}
        \label{subfig:RM-TL1-TL2-TL3}
    \end{subfigure}
    \hspace{-0.5cm}
    \begin{subfigure}{0.51\linewidth}
        \centering
        \includegraphics[width=0.91\linewidth]{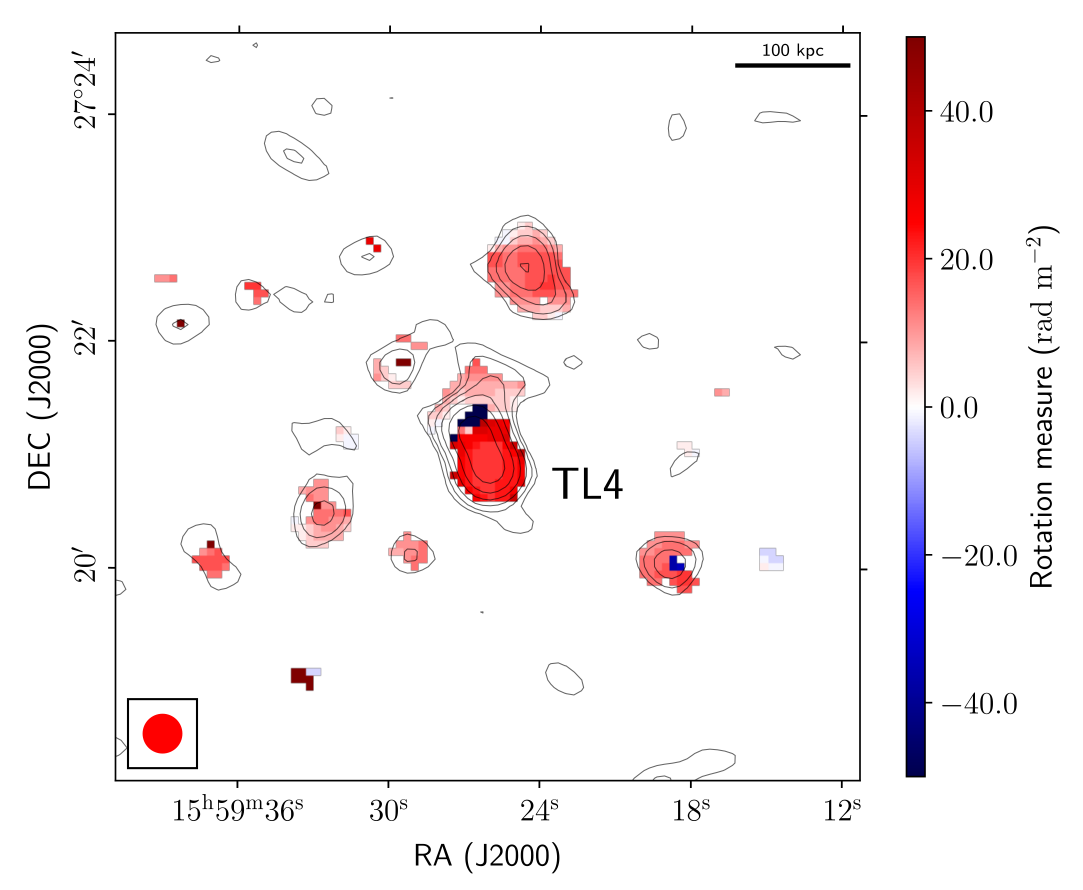}
        \caption{}
        \label{subfig:RM-TL4}
    \end{subfigure}
    \begin{subfigure}{0.51\linewidth}
        \hspace{-0.3cm}
        \centering
        \includegraphics[width=0.91\linewidth]{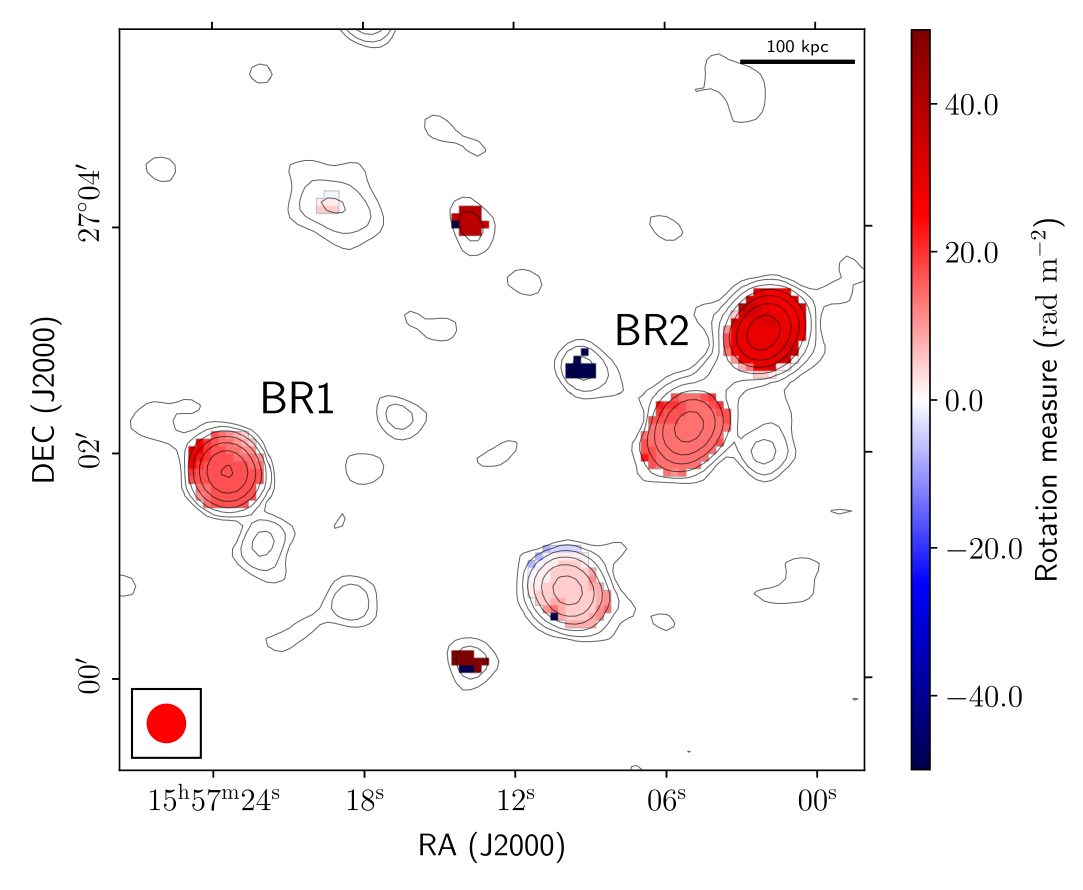} 
        \caption{}
        \label{subfig:RM-BR1-BR2}
    \end{subfigure}
    \hspace{-0.5cm}
    \begin{subfigure}{0.51\linewidth}
        \centering
        \includegraphics[width=0.91\linewidth]{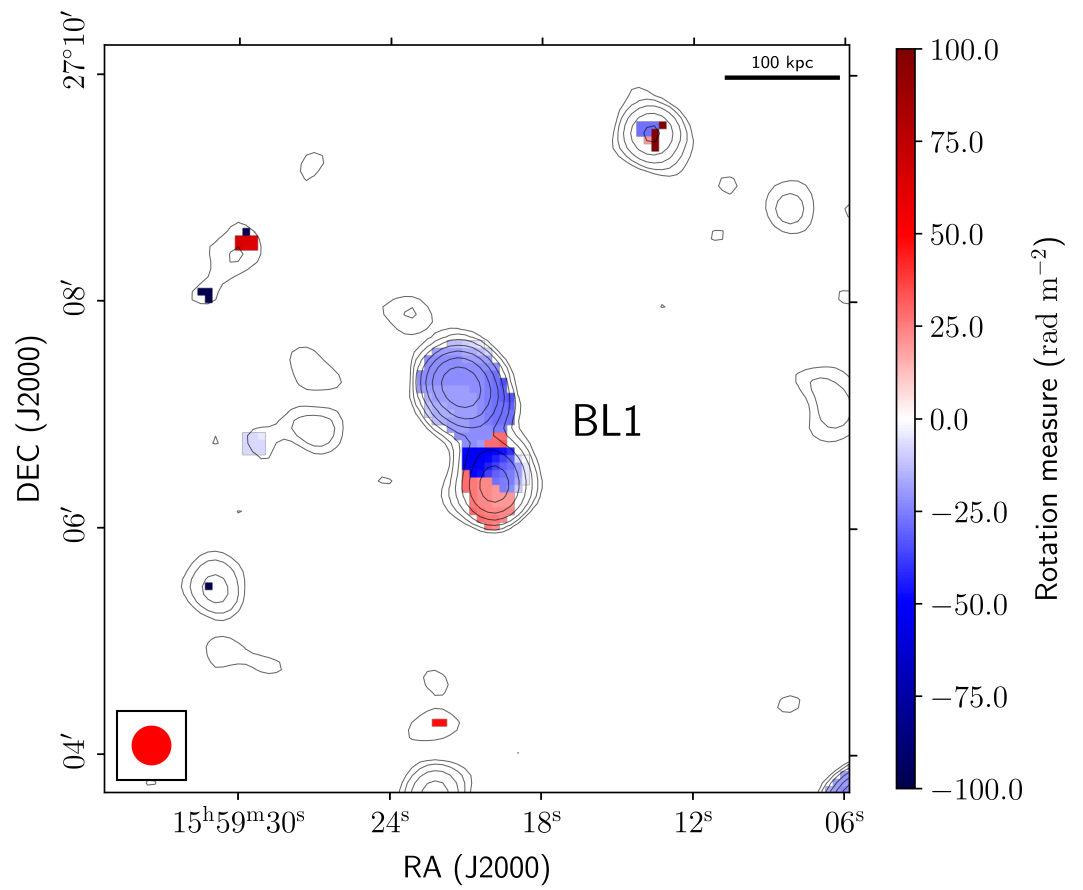} 
        \caption{}
        \label{subfig:RM-BL1}
    \end{subfigure}
    \begin{subfigure}{0.51\linewidth}
        \centering
        \includegraphics[width=0.91\linewidth]{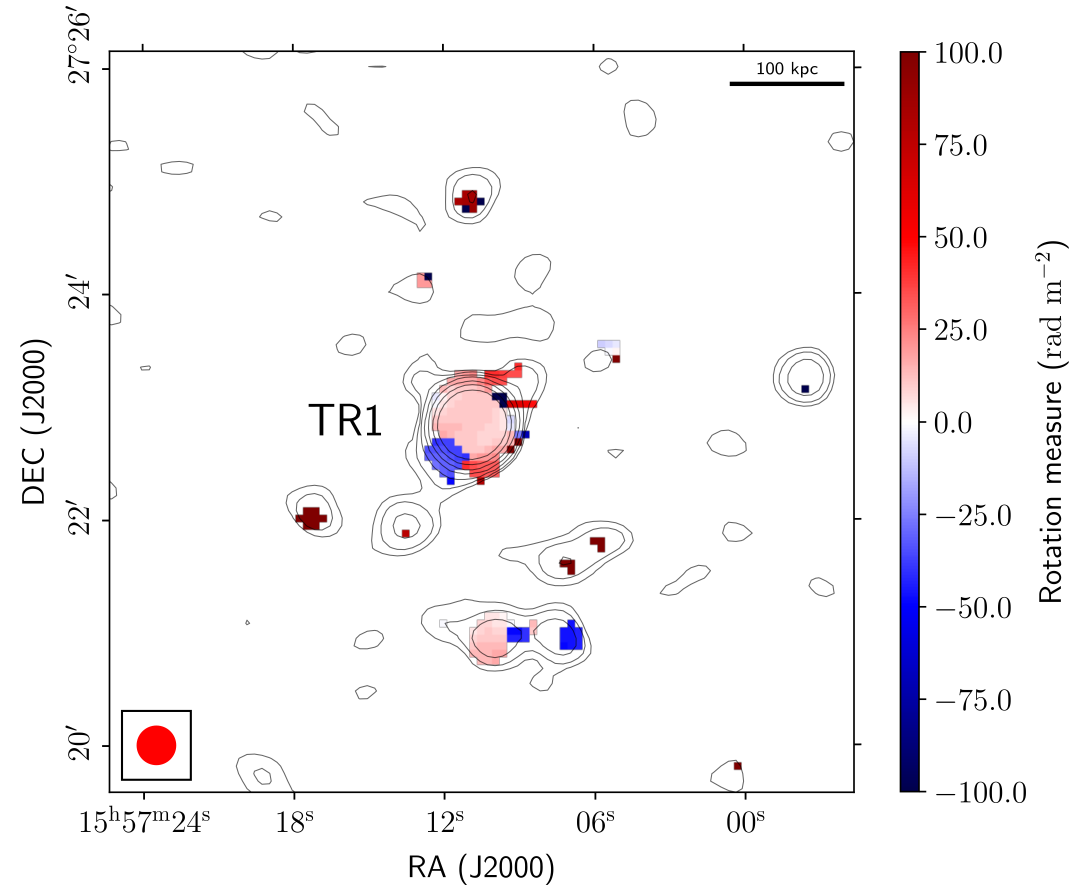} 
        \caption{}
        \label{subfig:RM-TR1}
    \end{subfigure}
    \label{fig:RM-sources-2}
\end{figure}

\begin{figure}[!htb]
    \caption{Zoom of the $\mathrm{F_p}$ map in correspondence of some sources analysed in this work with overlaid radio contours (1283 MHz), from $3\sigma_{\mathrm{I}}$ and scaling by a factor of 2. The $6\sigma_{\mathrm{QU}}$ and $3\sigma_{\mathrm{I}}$ detection thresholds were imposed in polarisation and in total intensity and only pixels above them are shown. Values were corrected for the Ricean bias. Panel (a): TL1, TL2 and TL3. Panel (b): TL4. Panel (c): BR1 and BR2. Panel (d): BL1. Panel (e): TR1. Spatial scales and resolution beam are reported on the edges of the images.}
    \begin{subfigure}{0.51\linewidth}
        \centering
        \includegraphics[width=0.91\linewidth]{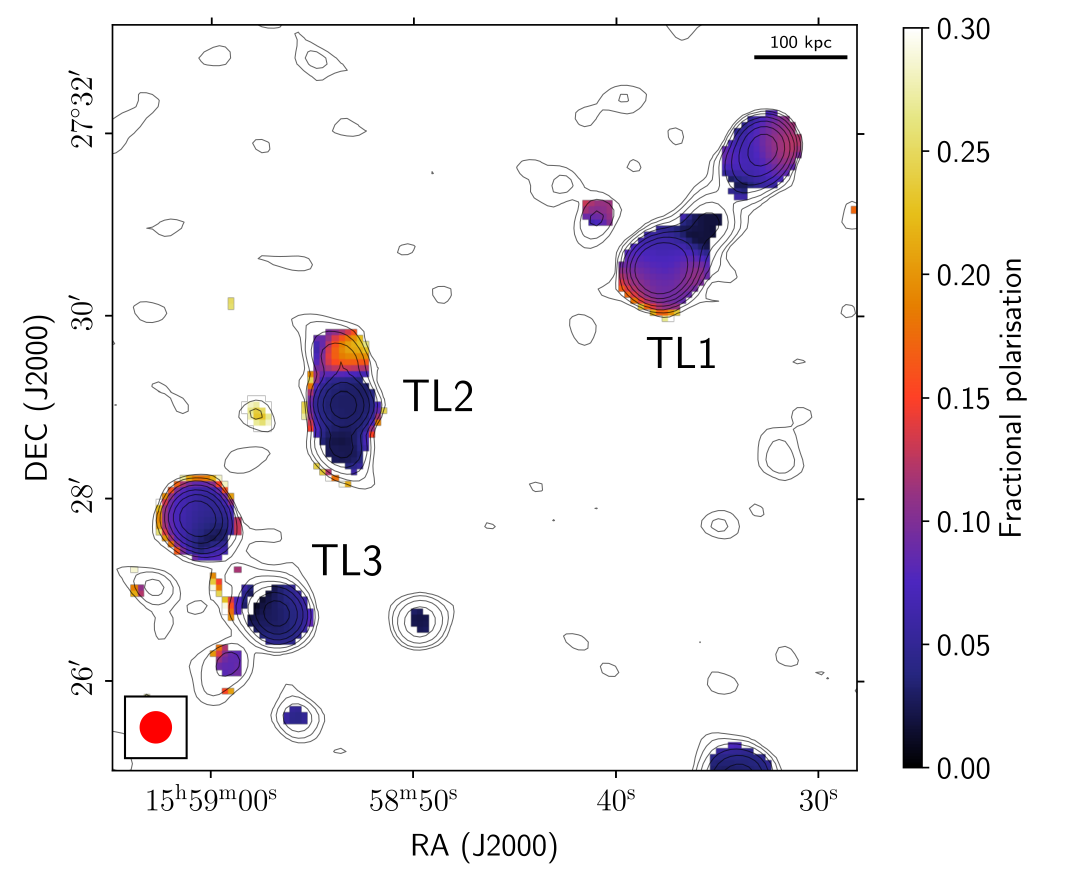}
        \caption{}
        \label{subfig:Fp-TL1-TL2-TL3}
    \end{subfigure}
    \hspace{-0.5cm}
    \begin{subfigure}{0.51\linewidth}
        \centering
        \includegraphics[width=0.91\linewidth]{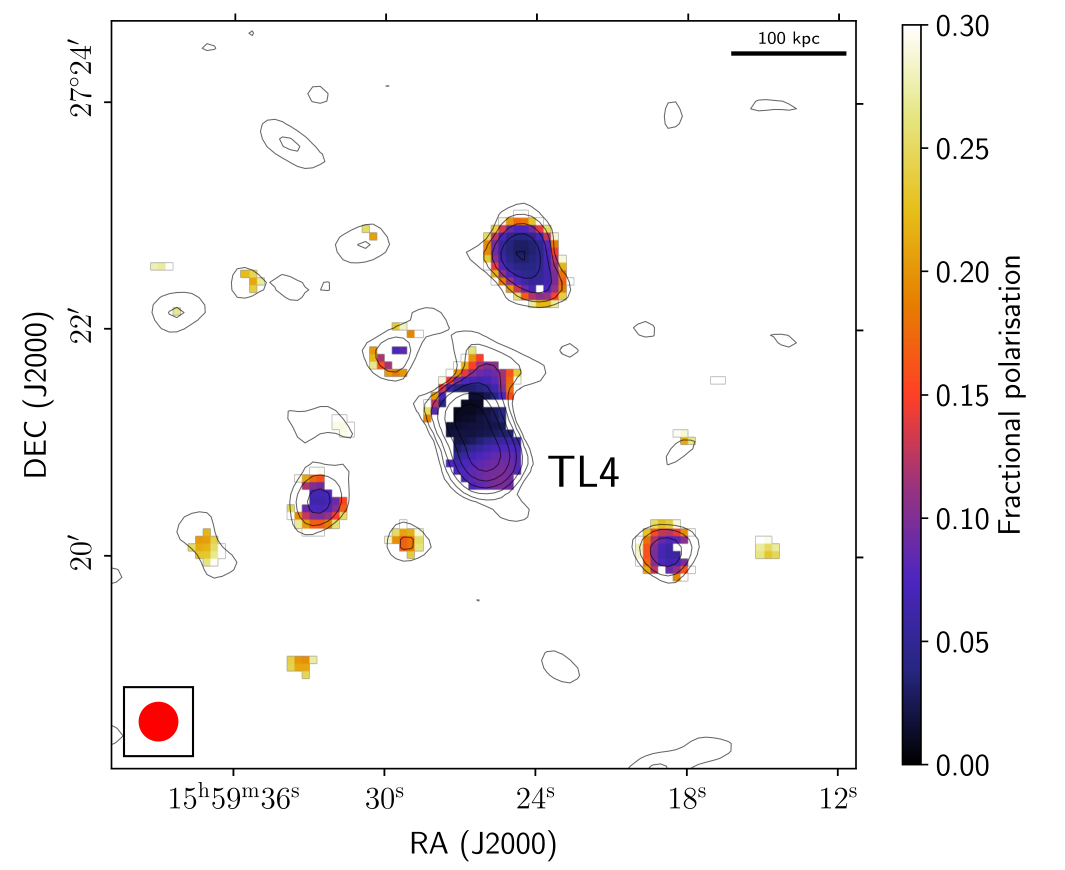}
        \caption{}
        \label{subfig:Fp-TL4}
    \end{subfigure}
    \begin{subfigure}{0.51\linewidth}
        \hspace{-0.3cm}
        \centering
        \includegraphics[width=0.91\linewidth]{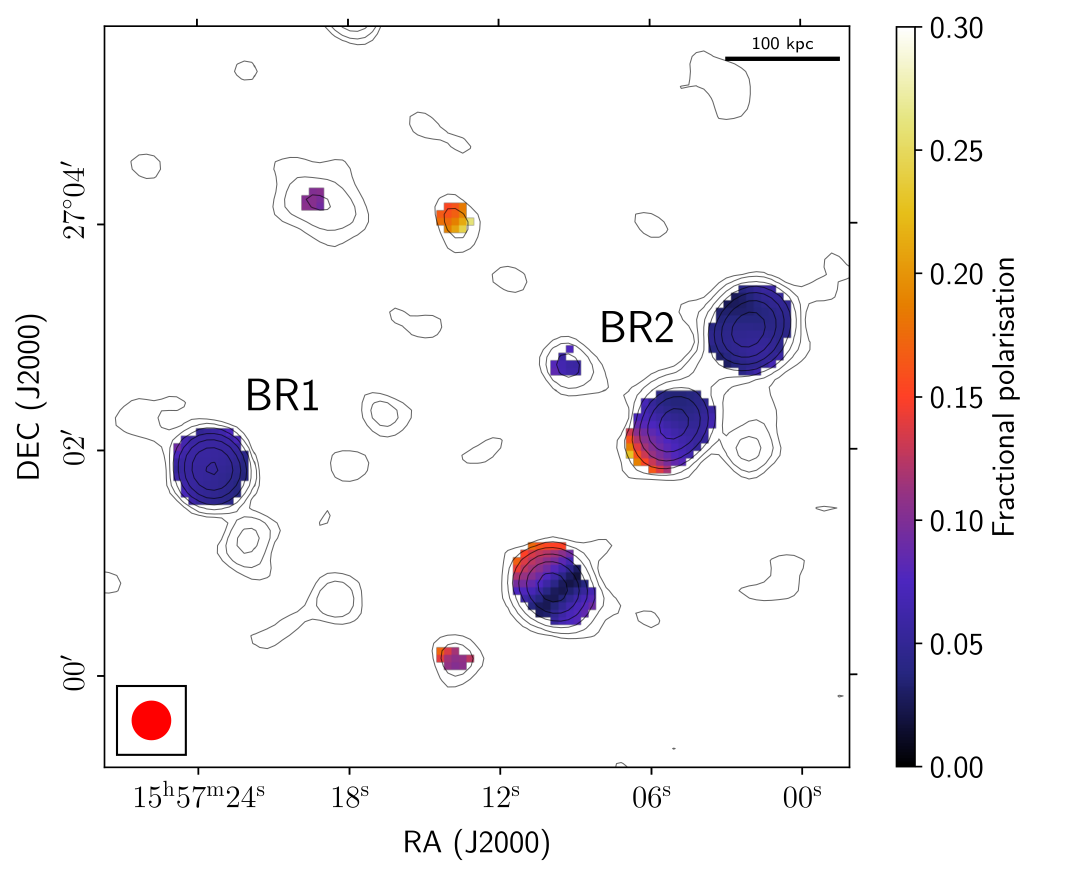} 
        \caption{}
        \label{subfig:Fp-BR1-BR2}
    \end{subfigure}
    \hspace{-0.5cm}
    \begin{subfigure}{0.51\linewidth}
        \centering
        \includegraphics[width=0.91\linewidth]{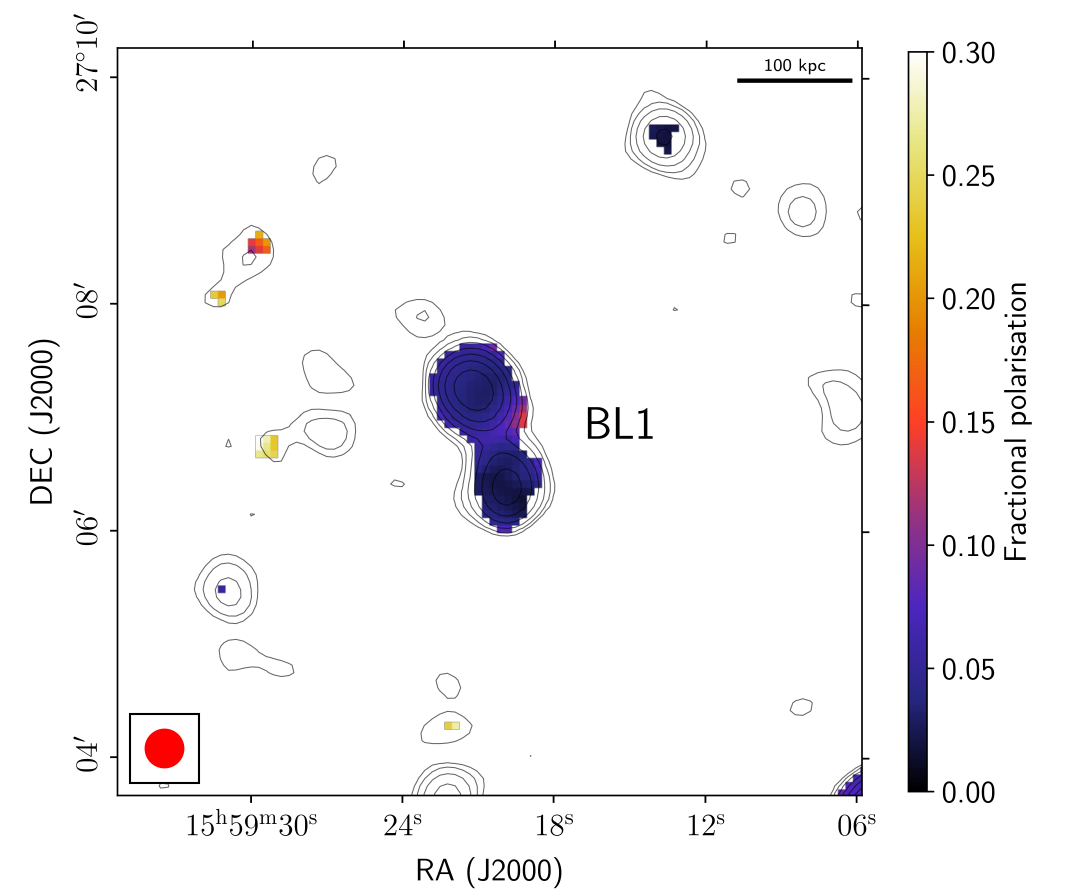} 
        \caption{}
        \label{subfig:Fp-BL1}
    \end{subfigure}
    \begin{subfigure}{0.51\linewidth}
        \hspace{-0.3cm}
        \centering
        \includegraphics[width=0.91\linewidth]{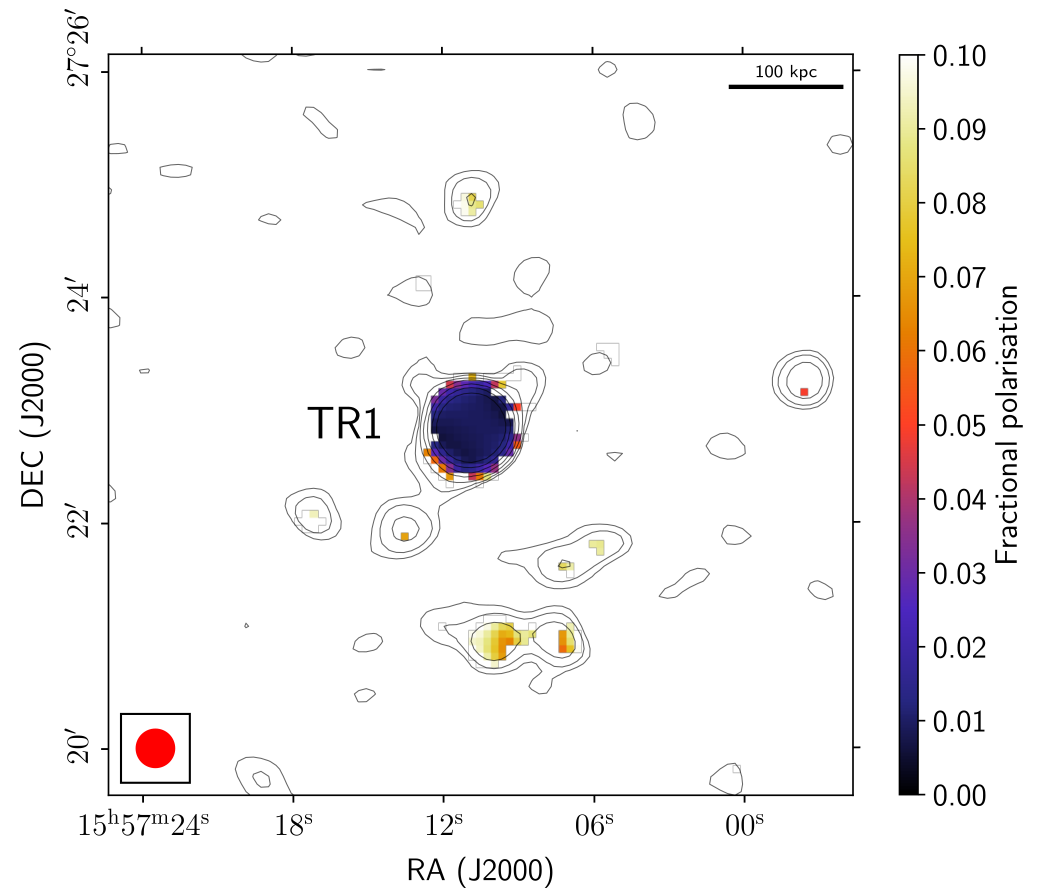} 
        \caption{}
        \label{subfig:Fp-TR1}
    \end{subfigure}
    \label{fig:Fp-sources-2}
\end{figure}

Here, we report the results obtained for other radio galaxies in the field. In particular, we show in Fig. \ref{fig:other_FDF} the comparison between the dirty FDF spectrum, the cleaned FDF spectrum after deconvolution and the clean components from the brightest polarised pixel of the source. All of them present a Faraday-complex spectrum due to multiple components along the same LOS, which with our resolution are very difficult to disentangle. In Figs. \ref{fig:RM-sources-2} and \ref{fig:Fp-sources-2} we present the RM maps and $\rm F_{p}$ maps of the unknown sources included in the radial profiles analysis. As mentioned in Sect. \ref{RM Synthesis Results}, the resolved sources seen in projection in the external regions show a more uniform distribution in the RM and lower values of $\rm |RM|$ with respect to what obtained at the centre, where, indeed, a stronger MF is expected. Moreover, the sources display typical percentages of fractional polarisation found for radio galaxies and not an evident depolarisation as for the central known objects.
\newpage
\newpage
\end{appendix}

\end{document}